\documentclass[a4paper,11pt]{article}
\pdfoutput=1 
\usepackage{jheppub} 

\usepackage[T1]{fontenc}

\title{\boldmath Rotating Hayward's regular black hole as particle accelerator}

\author[a]{Muhammed Amir}
\author[a,b,1]{and Sushant G. Ghosh\note{Corresponding author.}}

\affiliation[a]{Centre for Theoretical Physics,
 Jamia Millia Islamia,  New Delhi 110025, India}
\affiliation[b]{Astrophysics and Cosmology Research Unit,
 School of Mathematics, Statistics \\ and Computer Science,
 University of KwaZulu-Natal, Private Bag X54001,\\
 Durban 4000, South Africa}

\emailAdd{amirctp12@gmail.com}
\emailAdd{sghosh2@jmi.ac.in}

\abstract{Recently, Ban\~{a}dos, Silk and West (BSW) demonstrated that the
extremal Kerr black hole can act as a particle accelerator with
arbitrarily high center-of-mass energy ($E_{CM}$)  when the
collision takes place near the horizon. The rotating Hayward's
regular black hole, apart from Mass ($M$) and angular momentum
($a$), has a new parameter $g$ ($g>0$ is a constant) that provides a
deviation from  the Kerr black hole. We demonstrate that for each
$g$, with $M=1$, there exist critical $a_{E}$ and $r_{H}^{E}$, which
corresponds to a regular extremal black hole with degenerate
horizons, and $a_{E}$ decreases whereas $r_{H}^{E}$ increases with
increase in $g$. While $a<a_{E}$ describe a regular non-extremal
black hole with outer and inner horizons. We apply the BSW process to
the rotating Hayward's regular black hole, for different $g$, and
demonstrate numerically that the $E_{CM}$ diverges in the vicinity of
the horizon for the extremal cases thereby suggesting that a
rotating regular black hole can also act as a particle accelerator
and thus in turn provide a suitable framework for Plank-scale
physics. For a non-extremal case, there always exist a finite upper
bound for the $E_{CM}$, which increases with the deviation parameter $g$.}

\begin{document} 
\maketitle
\flushbottom

\section{Introduction}
\label{sec:intro}

Recently, Ban\~{a}dos, Silk and West (BSW) \cite{Banados:2009pr} have demonstrated that the collision of two particles falling from rest at infinity into the Kerr black hole \cite{Kerr:1963ud} can have an infinitely large center-of-mass energy ($E_{CM}$) close to the event horizon, if the black hole is maximally spinning, and one of the particle have critical angular momentum. This mechanism of particle acceleration by a black hole is called BSW mechanism, which is interesting from the viewpoint of theoretical physics because new physics is possible in the vicinity of the black holes at the Planck scale. It may enable us to explain the astrophysical phenomenon such as the gamma ray burst and the active galactic nuclei. Hence, the BSW mechanism about the collision of two particles near a rotating black hole has attracted much attention \cite{Berti:2009bk,Banados:2010kn,Jacobson:2009zg,thorne,Zaslavskii:2010aw,Wei:2010gq,Hussain:2012zza} (see also \cite{Harada:2014vka}, for  a review).  Subsequently, Lake \cite{Lake:2010bq} examined the $E_{CM}$ of the collision  at the inner horizon of the non-extremal Kerr black hole and found that the  $E_{CM}$ is finite. The BSW mechanism for the charged spinning black hole or the Kerr-Newman black hole \cite{ Wei:2010vca} was also addressed and shown that the  of collision of two uncharged particles falling freely from rest at infinity not only depends on the spin $ a $ but also on the charge $ Q $ of the black hole. The BSW effect for the Kerr-Taub-NUT BH was investigated in \cite{Liu:2010ja}, and around the four dimensional Kaluza-Klein extremal black hole in \cite{Mao:2010di}, and it resulted infinitely large $E_{CM}$ near the horizon of both rotating and non-rotating cases. Further,  the mechanism was generalized for charged particles moving in an electromagnetic field and for the braneworld black holes \cite{Zhu:2011ae}. Zaslavskii \cite{Zaslavskii:2012fh,Zaslavskii:2010pw,Zaslavskii:2012qy} elucidated the universal property of acceleration of particles for the rotating black holes and try to give a general explanation. Furthermore, Grib and Pavlov \cite{ Grib:2010xj,Grib:2012iq} found that the $E_{CM}$ can be infinite near the non-extremal Kerr black hole, if the multiple scattering of the colliding particles is included. The BSW mechanism was further extended to the case of two different massive colliding particles near the Kerr black hole \cite{Harada:2011xz} and also in the case of Kerr-Newman black hole \cite{Liu:2011wv}. It turns out that the divergence of the $E_{CM}$ of colliding particles is a phenomenon not only associated with black holes, but also with naked singularities \cite{Patil:2010nt,Patil:2011aw,Patil:2011ya,Patil:2011uf}.

On the other hand a spacetime singularity or a naked singularity is the final fate of continual gravitational collapse \cite{Joshi:2000fk}, and it is widely believed that a singularity must be removed by quantum gravity effects. However, we are far away from well defined quantum gravity, and hence much attention is devoted to the research on the properties and implications of classical black holes with a regular or non-singular center. In particular, an interesting proposal was made by Hayward \cite{Hayward:2005gi} for the formation and evaporation of a regular black hole based on the idea of Bardeen \cite{bardeen}, who proposed the first regular black hole. These black holes are solutions of modified Einstein's equation, yielding alteration to classical black holes, but near the center they behave like a de Sitter spacetime \cite{bardeen,Hayward:2005gi}. Over the past few years there has been an increasing interest in the study of rotating regular black holes \cite{Bambi:2013ufa,Toshmatov:2014nya}, which depend on the mass and  spin of the black hole, and on an additional deviation parameter that measure potential deviations from the Kerr metric, and includes the Kerr metric as the special case if this deviation parameter vanishes. Further, these regular black holes are very important as astrophysical black holes, like Cygnus X-1, although suppose to be like the Kerr black hole \cite{Bambi:2011mj,Bambi:2013qj}, but the actual nature of astrophysical black hole still need to be tested \cite{Bambi:2011mj,Bambi:2013qj}, and they may deviate from the Kerr black hole. More recently, the BSW  mechanism when applied to an extremal regular black holes \cite{Ghosh:2014mea,Pradhan:2014oaa}, also lead to divergence of the $E_{CM}$. The main purpose of this paper is to study the collision of two particles with equal rest masses in the background of the rotating Hayward's regular black hole and to see what effect the deviation parameter $g$ makes on the $E_{CM}$. It may be mentioned that the rotating Hayward's regular black hole is a prototype of a non-Kerr black hole with additional parameter $g$ apart from $M$ and $a$, which looks like the Kerr black hole with different spin \cite{Bambi:2013ufa}, may be a suitable candidate for an astrophysical black hole. Further, if observation demands vanishing deviation parameter, the compact object may be regarded as the Kerr black hole or a non-Kerr black hole otherwise. It turns out that observation may permit both these cases. We also study the horizon structure of the rotating Hayward's regular black hole, explicitly show the effect of the deviation parameter $g$. Further, our results go over to that of the Kerr black hole when parameter $g$ vanish, and to the nonrotating Hayward's regular black hole when $a=0$.

Further, there are several questions that motivate our analysis: how does the deviation term $g$ affect the BSW mechanism?  What is the horizon structure in the presence of term $g$? Whether such solutions lead to some important outcome? Do the calculation of the $E_{CM}$ has departed from the Kerr black hole? As we will see, these regular solutions do have several interesting features and consequences on the BSW mechanism. The paper is structured as follows. In the next section, we review the rotating Hayward's regular black hole, and discuss in detail its horizon structure. The calculation of the $E_{CM}$ for two colliding particles coming from rest at infinity in the background of the rotating Hayward's regular black hole is the subject of section IV. The required equation of motion, study of effective potential and calculation of the range of angular momentum is the subject of section III. We conclude the paper in section V.  We have used units which fix the speed of light and the gravitational constant via $8\pi G = c^4 = 1$.

\section{Rotating Hayward's regular black hole}
The aim of this paper is to demonstrate that the rotating Hayward's regular black hole can act as a particle accelerators. The metric of the rotating Hayward's regular solution, in the Boyer-Lindquist coordinates, which is equivalent to the Kerr metric \cite{Kerr:1963ud}, reads \cite{Bambi:2013ufa}
\begin{eqnarray}\label{hay}
d{s}^2 &=& -\left(1-\frac{2mr}{\Sigma}\right)dt^2 -\frac{4amr \sin^2 \theta}{\Sigma}dtd\phi +\frac{\Sigma}{\Delta}dr^2
\nonumber \\
&+& \Sigma d\theta^2 +\left(r^2+a^2+\frac{2a^2mr \sin^2 \theta}{\Sigma}\right)\sin^2 \theta d\phi^2,
\end{eqnarray}
where
\begin{eqnarray}
\Sigma =r^2 + a^2\cos^2\theta,\;\;\;\;\; \Delta = r^2-2mr+a^2,
\end{eqnarray}
with mass function is replaced by $m_{\alpha, \beta}(r,\theta)$ \cite{Bambi:2013ufa}, given by
\begin{eqnarray}\label{m}
m\rightarrow m_{\alpha, \beta}(r,\theta) = M\frac{r^{3+\alpha}
\Sigma^{-\alpha/2}}{r^{3+\alpha} \Sigma^{-\alpha/2}+g^3r^{\beta}
\Sigma^{-\beta/2}},
\end{eqnarray}
where, in general, $ m_{\alpha, \beta}(r,\theta) $ is a function of
$r$ and $\theta$ and also characterized by the two real numbers
$\alpha$ and $\beta$. The constants $\alpha$, $\beta$ are two real
numbers, $g$ is a positive constant, $a=J/M$ is angular momentum per
unit mass, and $M$ is the mass of the black hole. Thus, the rotating
Hayward's regular metric can be seen as a prototype of a large
class of a non-Kerr black hole metrics, in which the metric tensor, in
Boyer-Lindquist coordinates, has the same expression of the Kerr one
with $ m $ replaced by a mass function $ m_{\alpha, \beta}(r,\theta)
$ with $g$ gives deviation from the standard Kerr solution
\cite{Kerr:1963ud} and one recovers the Kerr solution in the limit
$g\rightarrow 0$. Further, in addition, if the rotational parameter
$a=0$, we get the Schwarzchild solution. The metric (\ref{hay}) is
regular everywhere, including at $r=0$ for $g \neq 0$, which can be
checked by the behavior of Ricci scalar and the Kretschman scalar. In
fact, the curvature invariants are regular everywhere, including at
$r=0$, where they remarkably zero \cite{Bambi:2013ufa}.

\begin{figure}[tbp]
\centering 
\includegraphics[width=.45\textwidth]{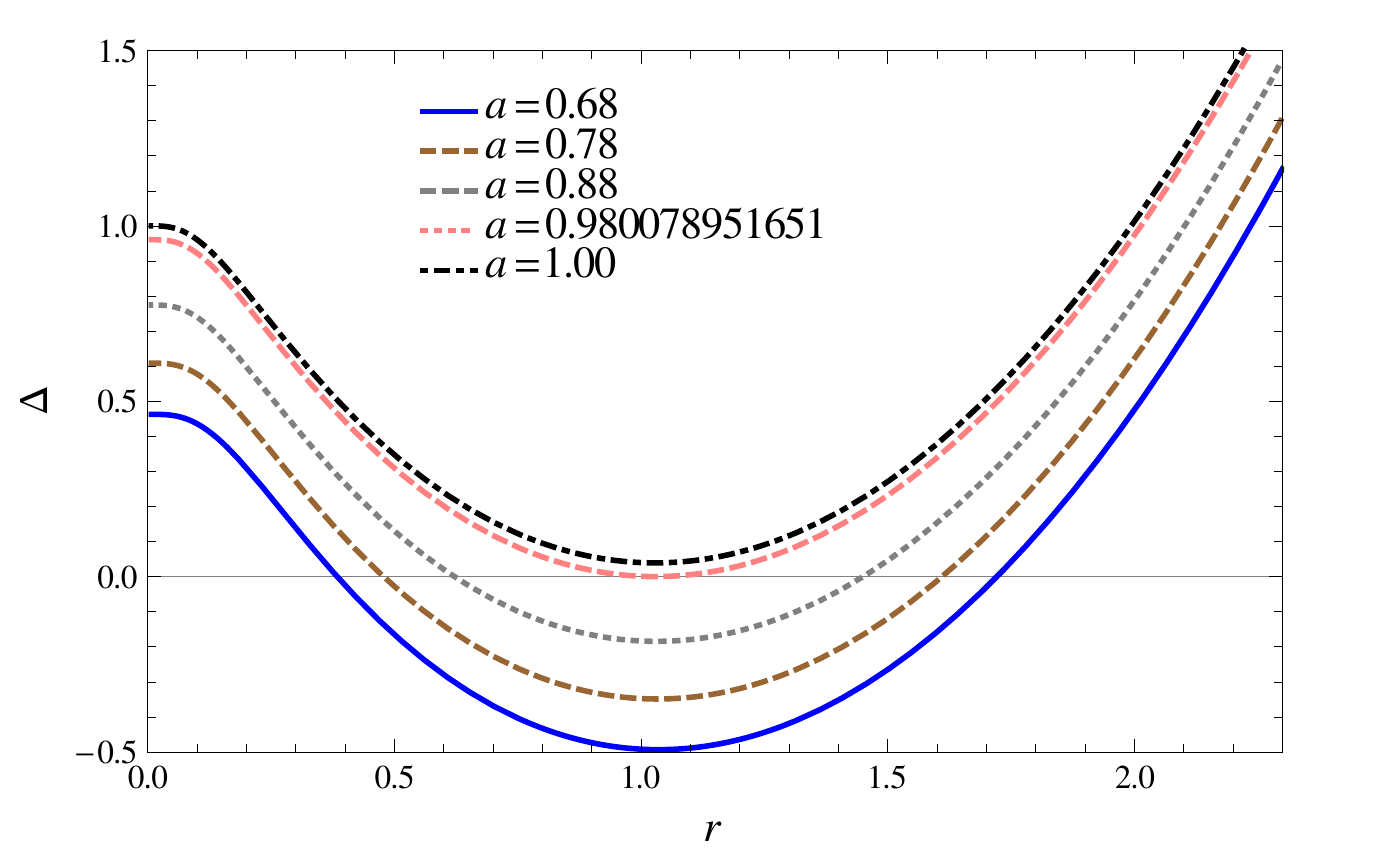}
\includegraphics[width=.45\textwidth]{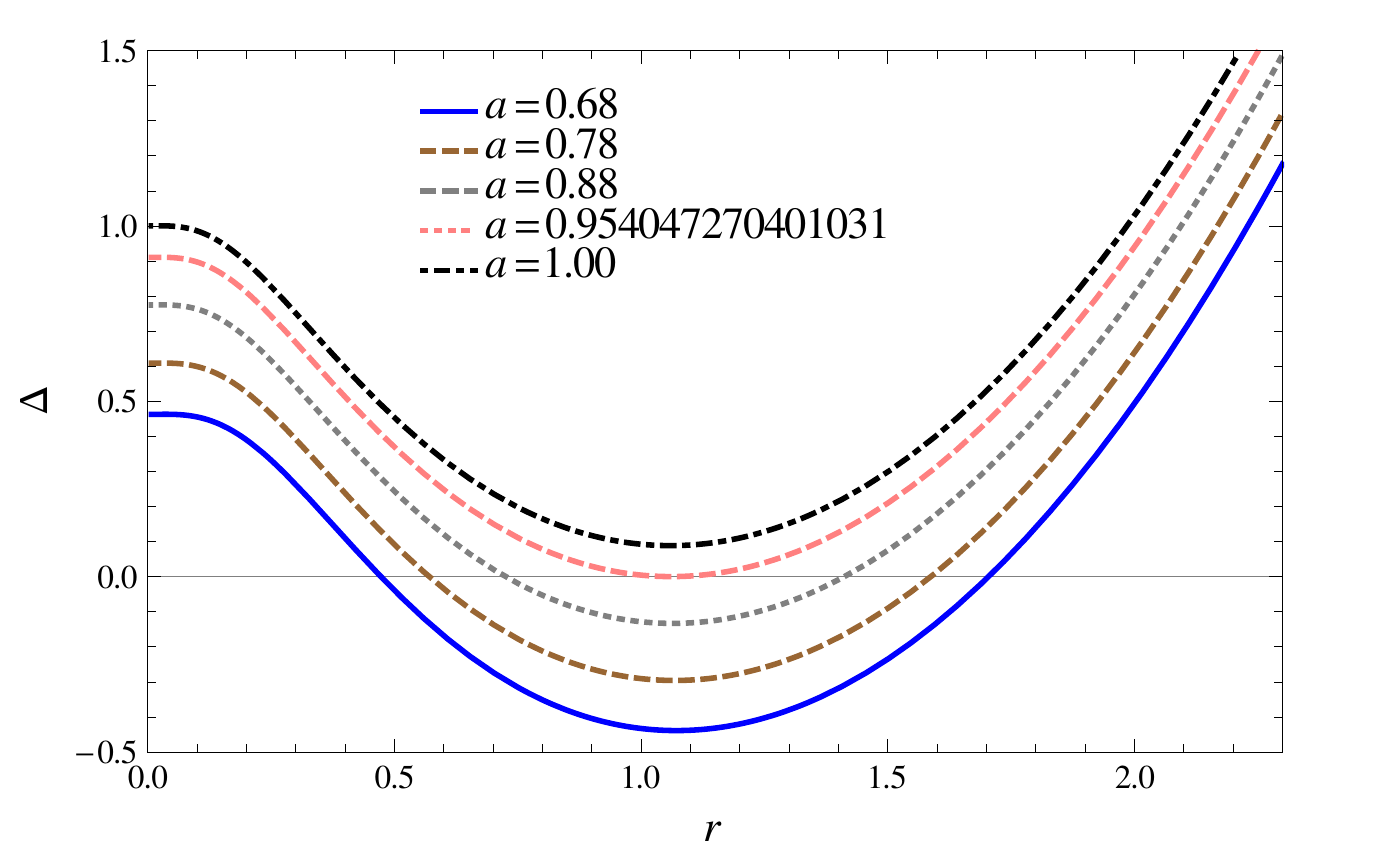}
\includegraphics[width=.45\textwidth]{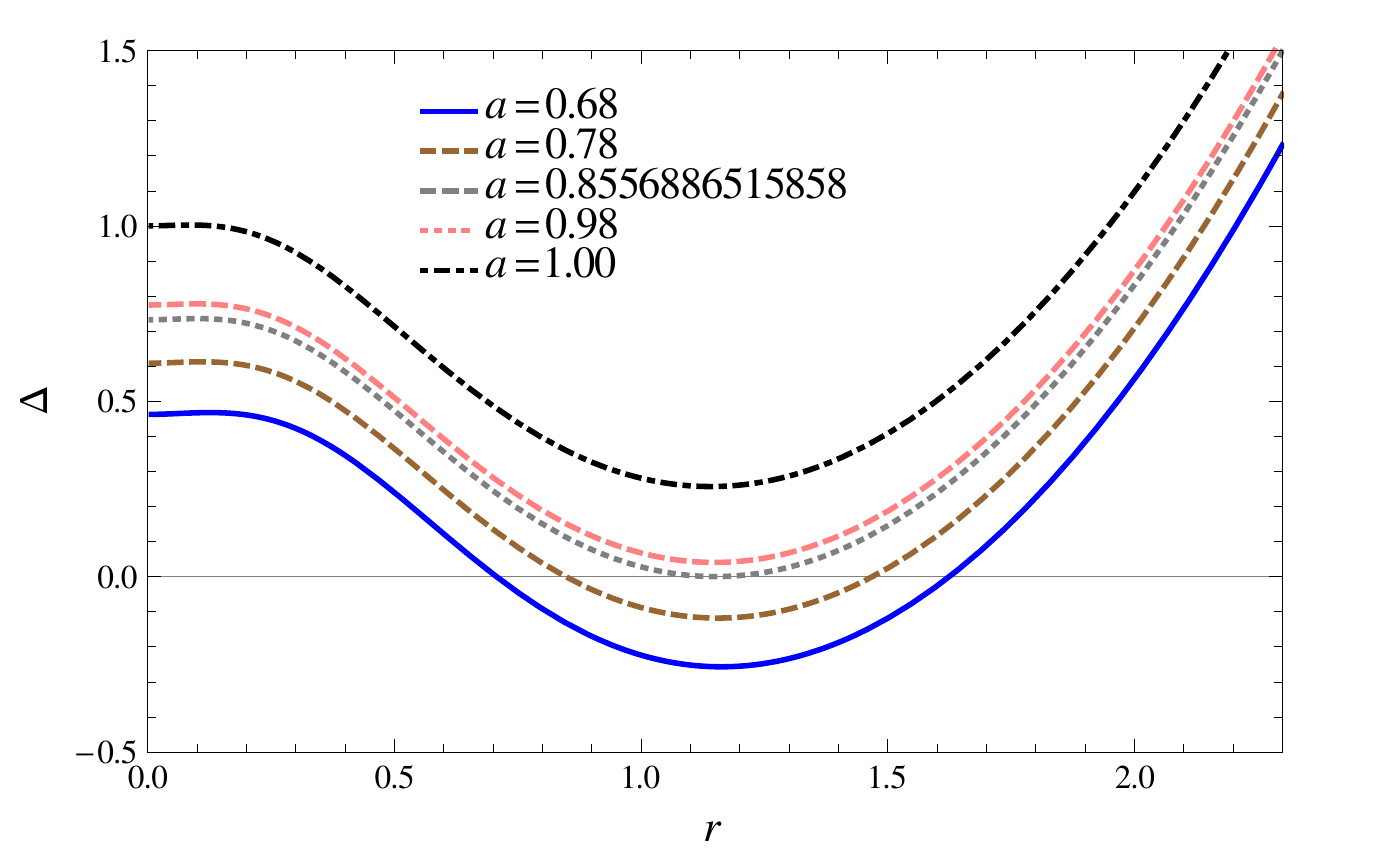}
\includegraphics[width=.45\textwidth]{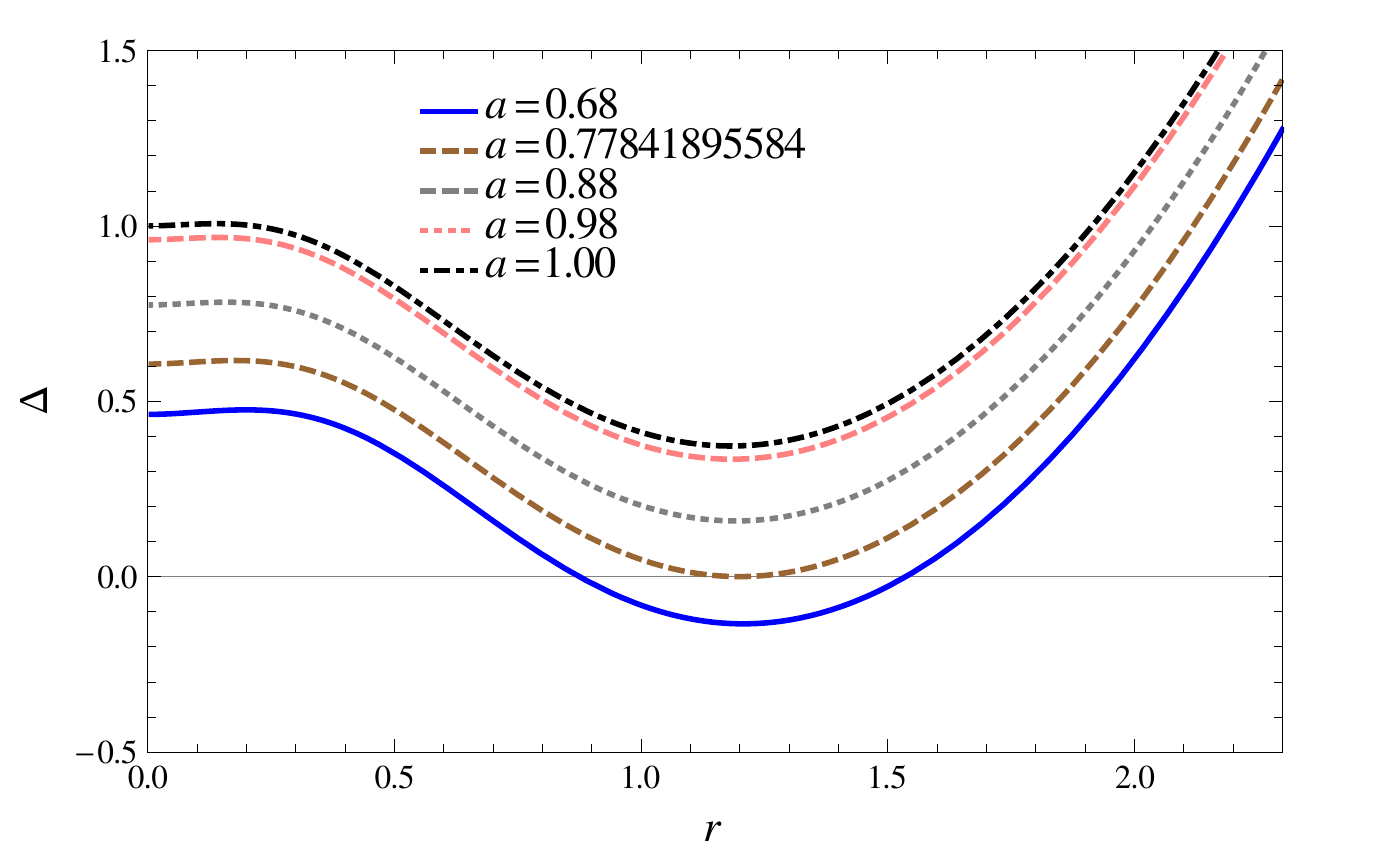}
\caption{\label{fig1} Plots showing the behavior of $\Delta$ vs $r$ for $\alpha=1$, $\beta=2$, $\theta=\pi/6$ and different values of $a$. Top: For $g=0.3$ (left), and $g=0.4$ (right). Bottom: For $g=0.6$ (left), and $g=0.7$ (right).}
\end{figure}

\begin{figure}[tbp]
\centering 
\includegraphics[width=.45\textwidth]{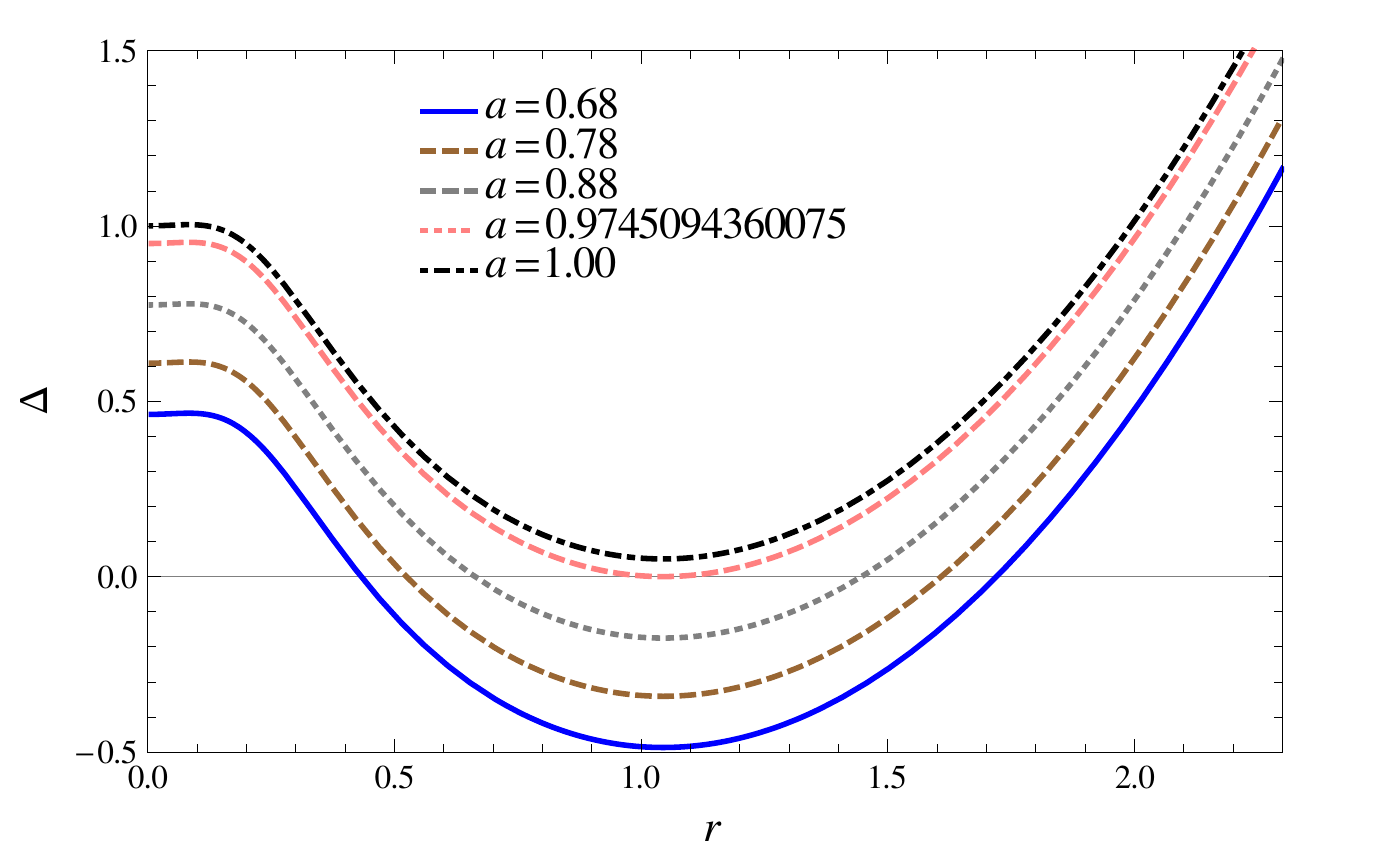}
\includegraphics[width=.45\textwidth]{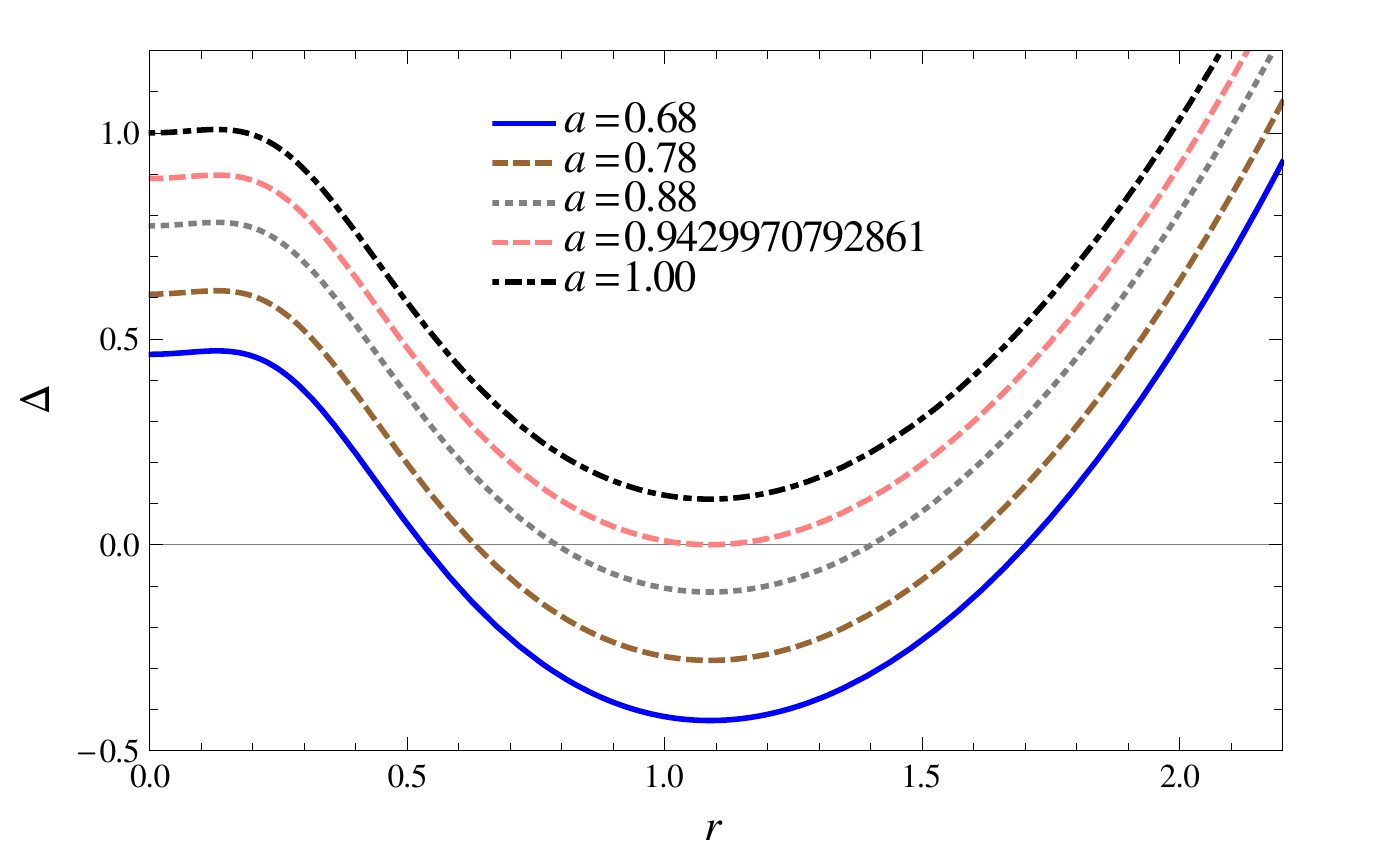}
\includegraphics[width=.45\textwidth]{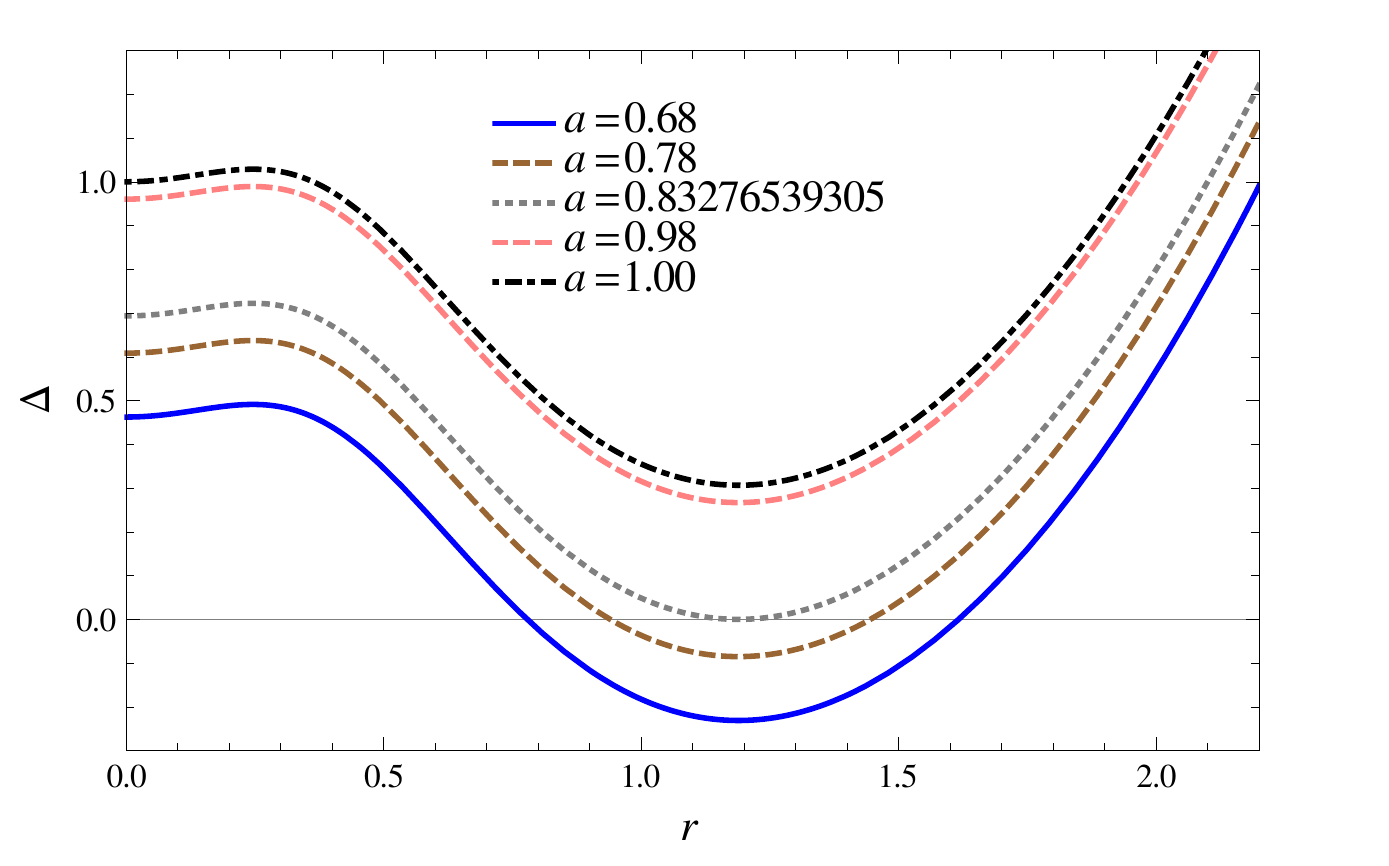}
\includegraphics[width=.45\textwidth]{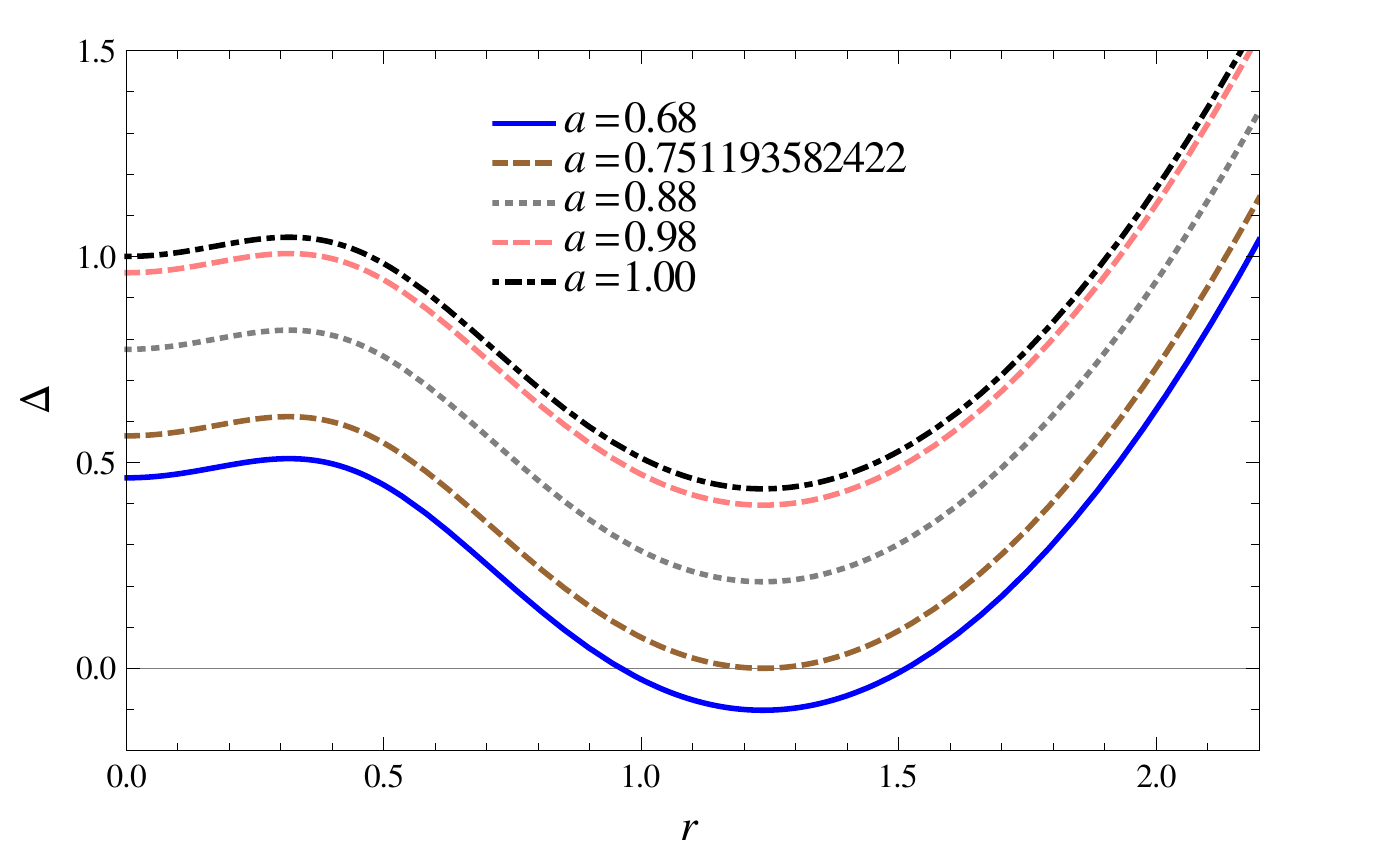}
\caption{\label{fig2} Plots showing the behavior of $\Delta$ vs $r$ for $\alpha=\beta=0$ and different values of $a$. Top: For $g=0.3$ (left), and $g=0.4$ (right). Bottom: For $g=0.6$ (left), and $g=0.7$ (right).}
\end{figure}

\begin{figure}[tbp]
\centering 
\includegraphics[width=.45\textwidth]{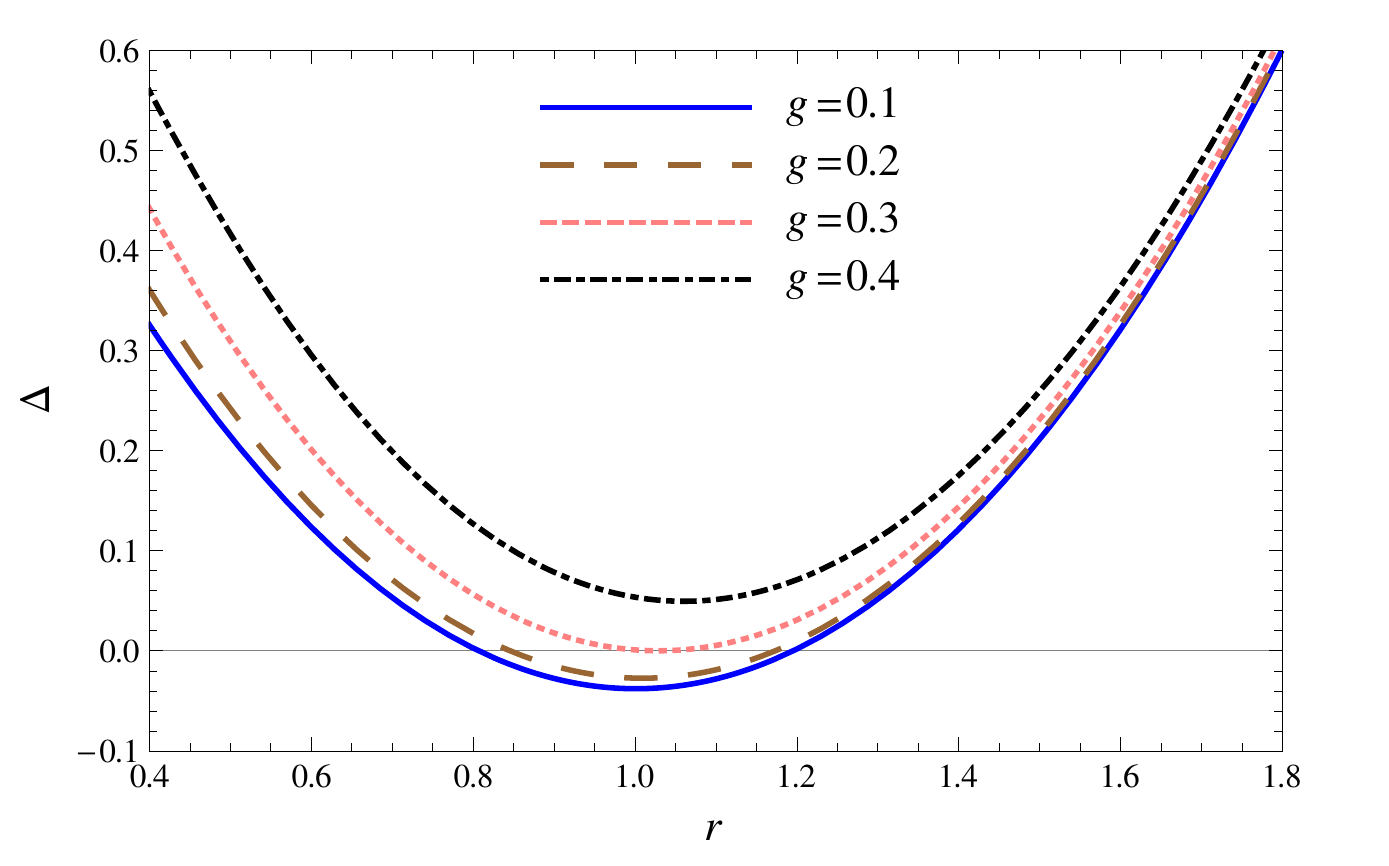}
\includegraphics[width=.45\textwidth]{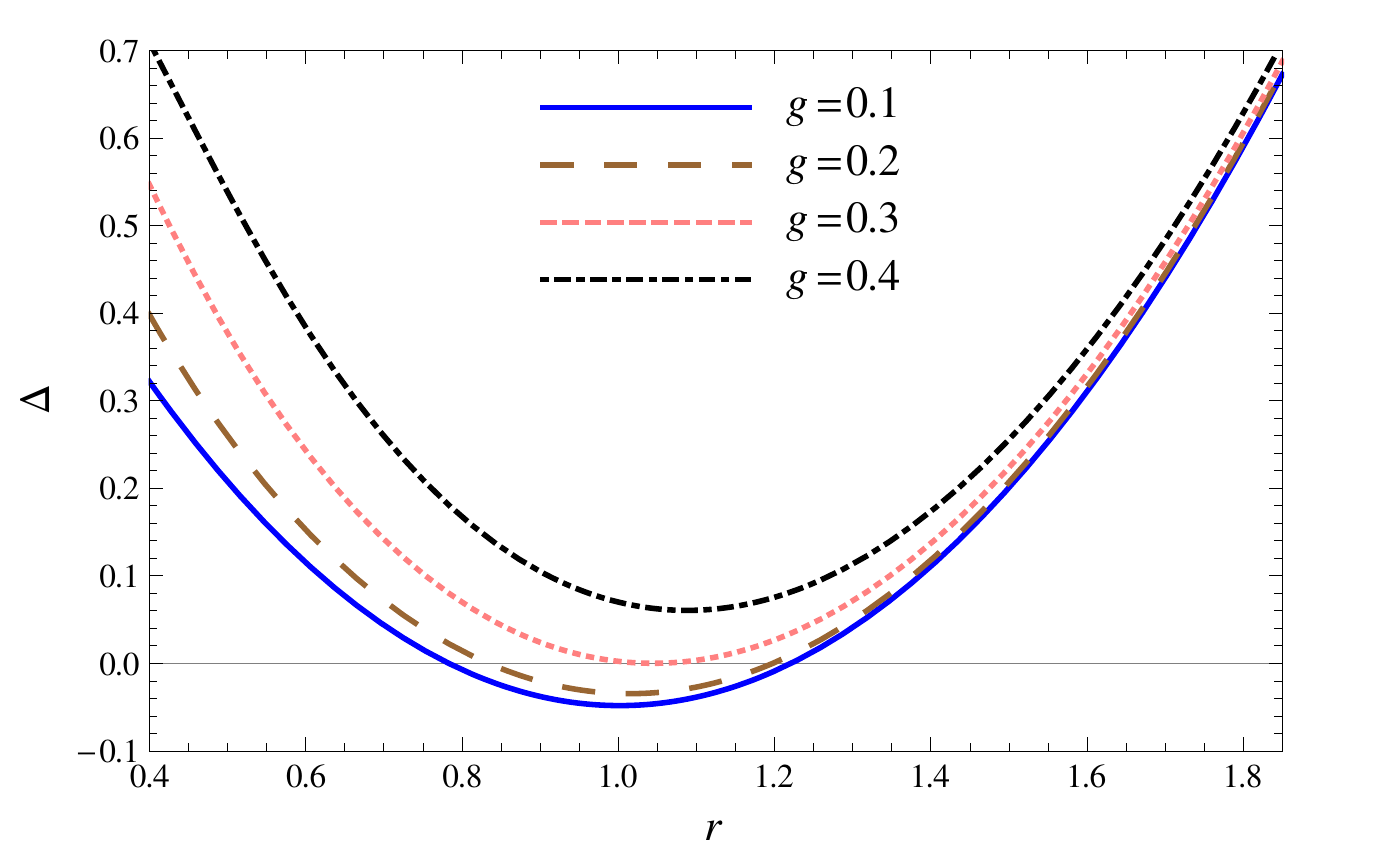}
\caption{\label{fig3} Plots showing the behavior of $\Delta$ vs $r$. (Left) For $\alpha=1$, $\beta=2$, $\theta=\pi/6$, $a=a_{E}=0.980078951651$ and different values of $g$. (Right) For $\alpha=\beta=0$, $a=a_{E}=0.9745094360075$ and different values of $g$.}
\end{figure}

\begin{figure}[tbp]
\centering 
\includegraphics[width=.45\textwidth]{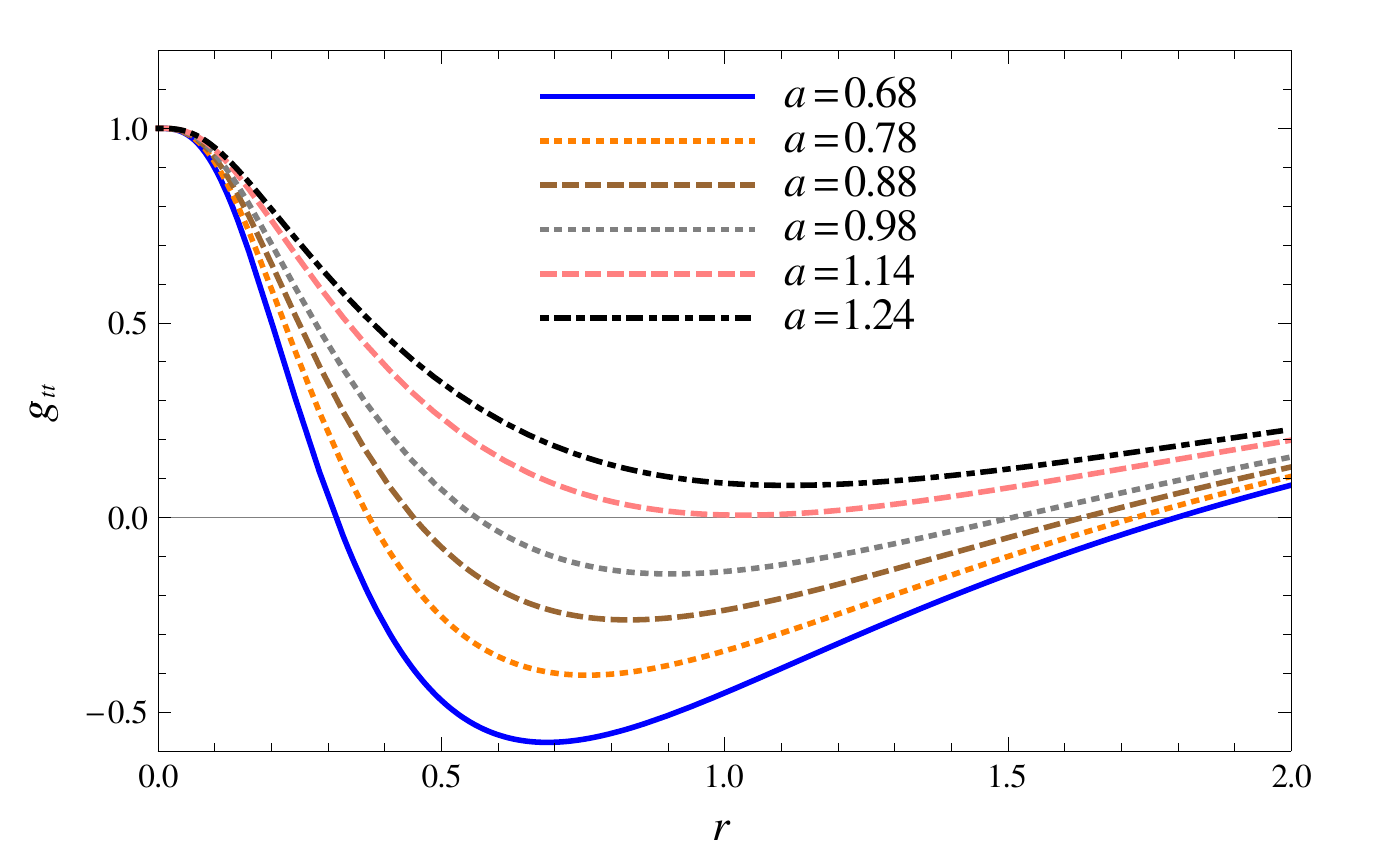}
\includegraphics[width=.45\textwidth]{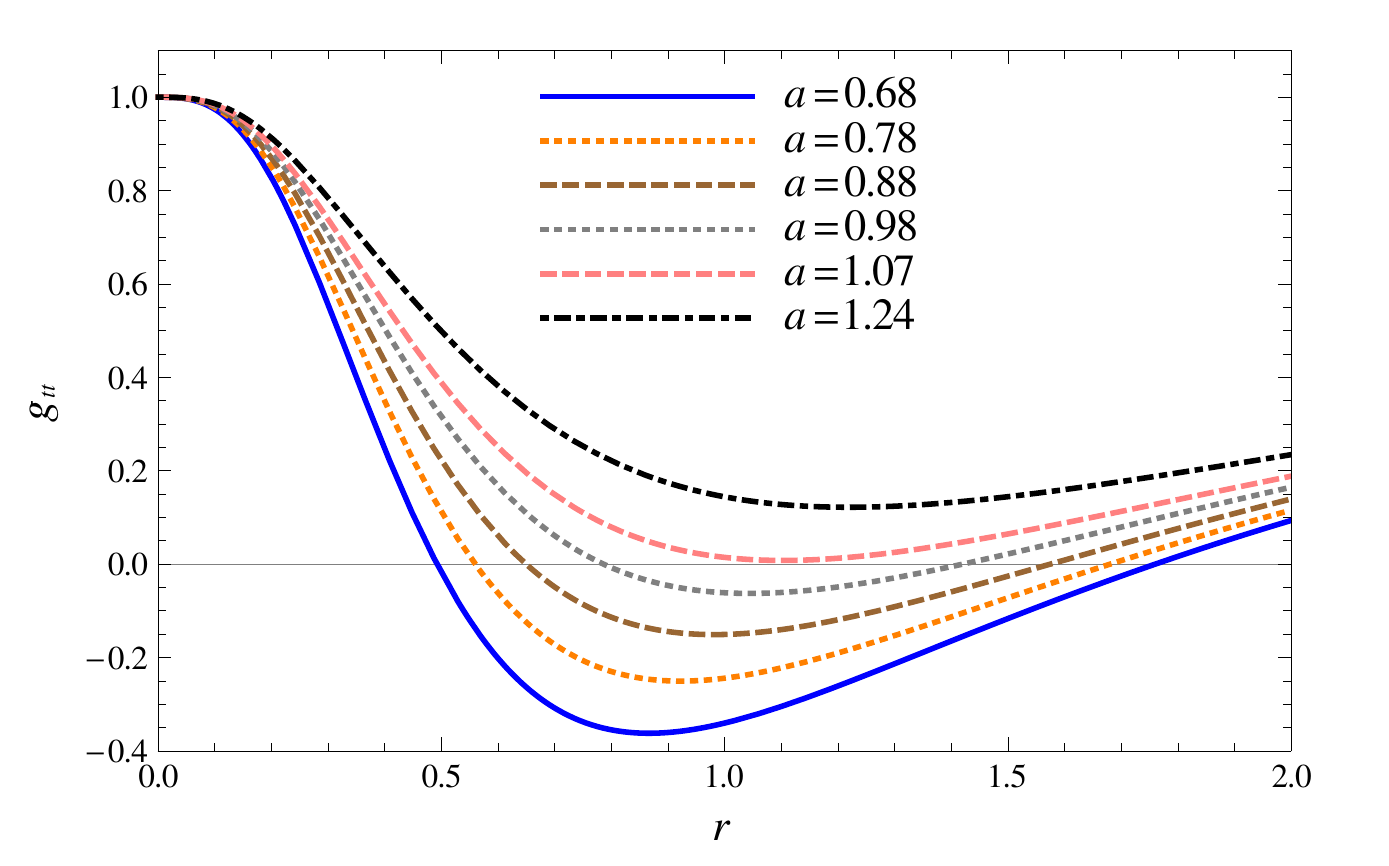}
\includegraphics[width=.45\textwidth]{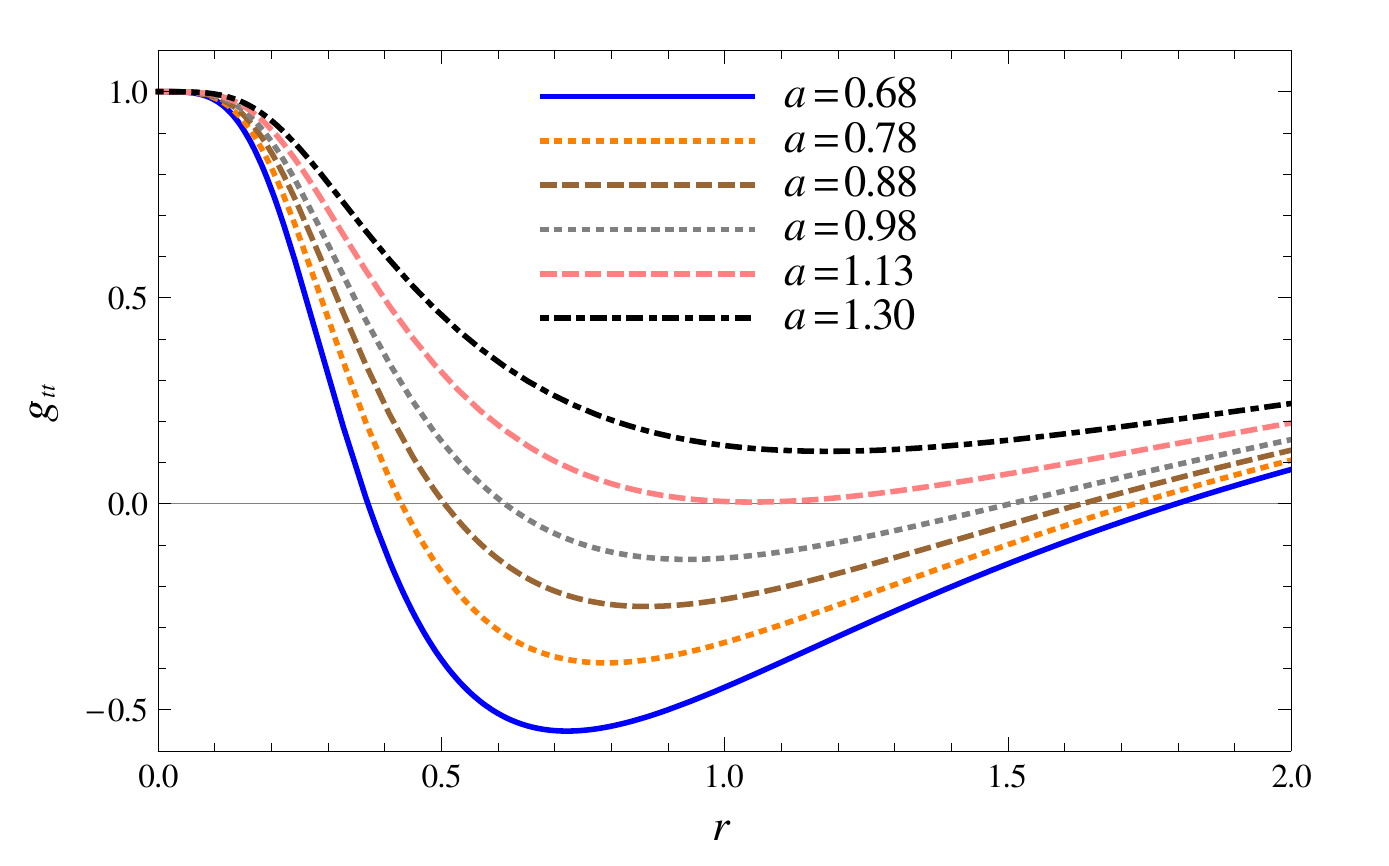}
\includegraphics[width=.45\textwidth]{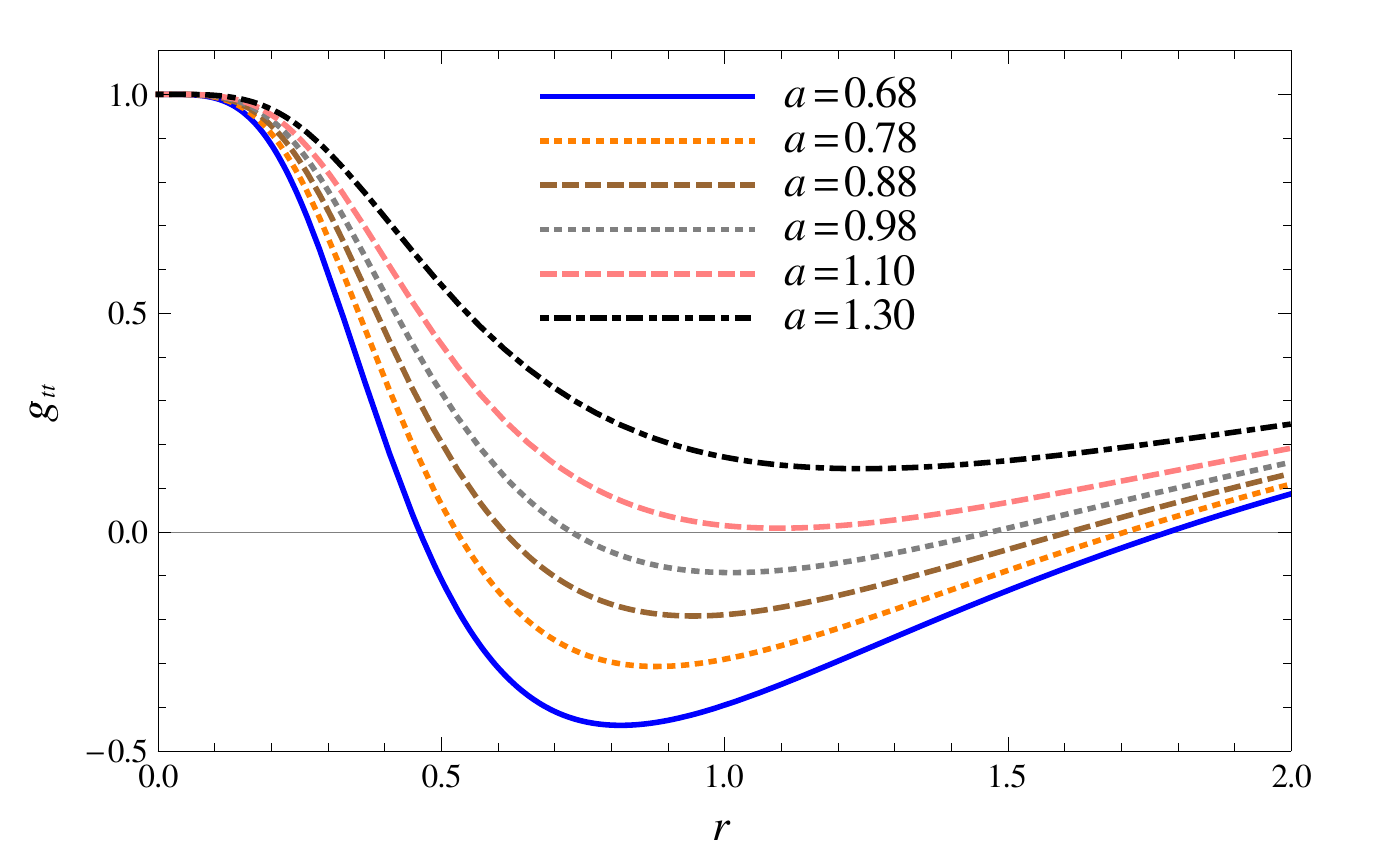}
\caption{\label{fig4} Plots showing the behavior of $g_{tt}$ vs $r$ for different values of $a$. Top: For $\alpha=1$, $\beta=2$, $\theta=\pi/6$, $g=0.3$ (left), and $g=0.4$ (right). Bottom:  For $\alpha=0$, $\beta=0$, $g=0.3$ (left), and $g=0.4$ (right).}
\end{figure}

\begin{figure}[tbp]
\centering 
\includegraphics[width=.3\textwidth]{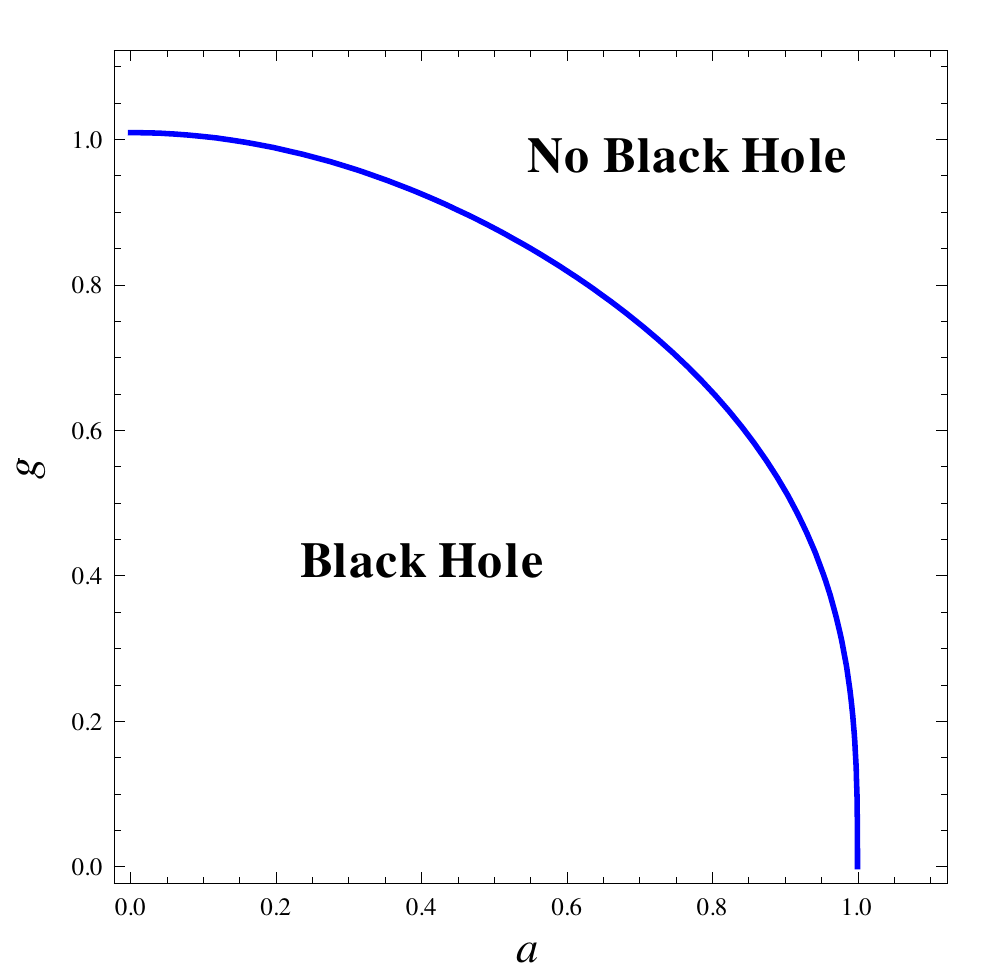}
\includegraphics[width=.3\textwidth]{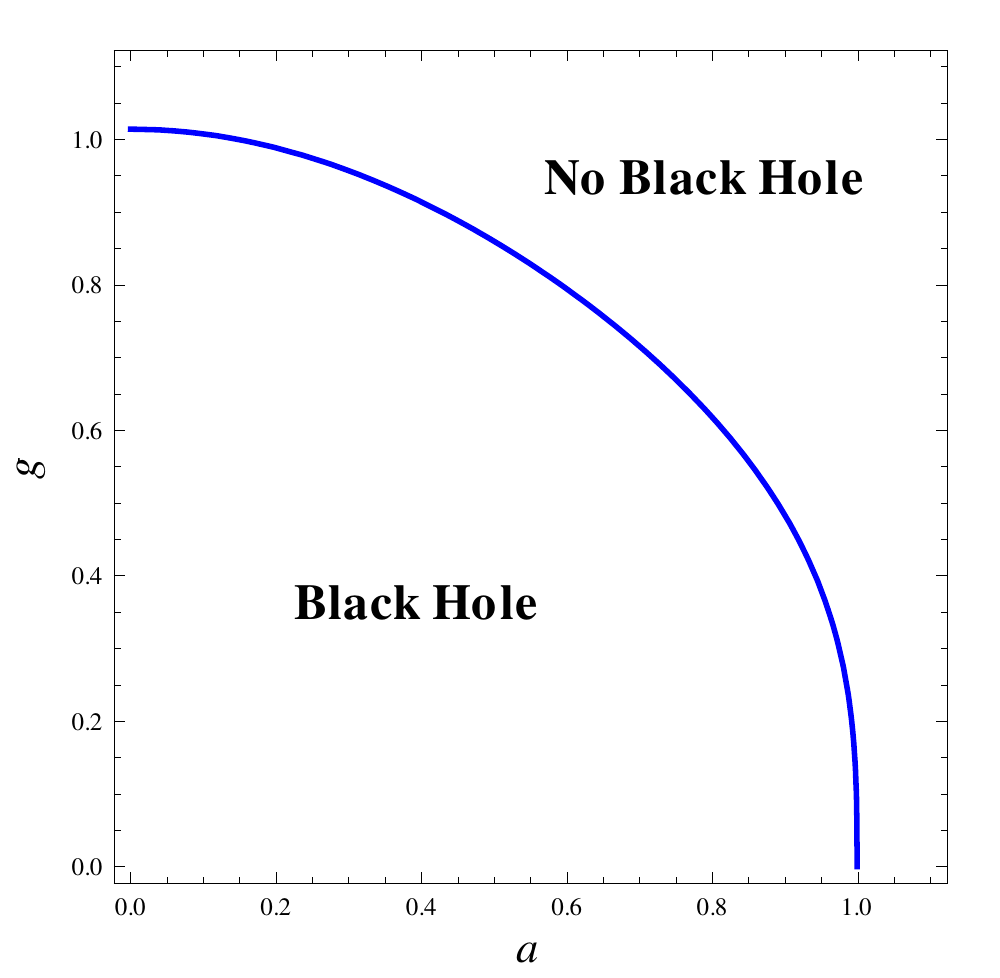}
\caption{\label{fig5} Plots showing the behavior of $g$ vs $a$. (Left) For $\alpha=1$, $\beta=2$ and $\theta=\pi/6$. (Right) For For $\alpha=\beta=0$.}
\end{figure}

The rotating Hayward's spacetime is stationary and axisymmetric with
Killing vector $\frac{\partial}{\partial t}$ and
$\frac{\partial}{\partial \phi}$. However, like the Kerr metric, the
Hayward's rotating metric (\ref{hay}) is also singular at
$\Delta=0$. The metric (\ref{hay}) generically must have two
horizons, viz., the Cauchy horizon and the event horizon. The
surface of no return is known as the event horizon. The zeros of
$\Delta=0$ gives the horizons of the black hole, i.e., the roots of
the following equation
\begin{eqnarray}\label{eh}
r^2(r^{3+\alpha}\Sigma^{-\alpha/2}+g^3r^{\beta}\Sigma^{-\beta/2})-2Mr^{4+\alpha}\Sigma^{-\alpha/2}\nonumber \\
+a^2 (r^{3+\alpha}\Sigma^{-\alpha/2}+g^3r^{\beta}\Sigma^{-\beta/2})&=&0, \nonumber \\
\end{eqnarray}
which depends on $a$, $g$, and $\theta$, and which is different from
the Kerr black hole. In the equatorial plane ($\theta=\pi/2$), it
turns out that mass function $ m_{\alpha, \beta}(r,\theta) $  is
independent of the parameters $\alpha$ and $\beta$, and so is
Eq.~(\ref{eh}), which is also the case when both $\alpha = \beta
=0.$  Thus, the two cases $\theta=\pi/2$, and $\alpha = \beta =0$
(for any $\theta$), will lead to the same mass function. Hence, the two tables are identical for these cases.

\begin{table}
    \begin{center}
        \caption{The event horizon ($r^{+}_{H}$) and the Cauchy horizon ($r^{-}_{H}$) of the black hole and their difference  $\delta^{g}=r^{+}_{H}-r^{-}_{H}$ for $\alpha=1, \beta=2$ for different values of spin $a$, with constant $g$.}\label{table1}
        \begin{tabular}{l l l l | l l l}
            \hline \hline
            &\multicolumn{3}{c}{$g=0.3$}  &  \multicolumn{3}{c}{$g=0.4$}\\
            \hline
$a$ & $r^{+}_{H}$ & $r^{-}_{H}$ & $\delta^{0.3}$ & $r^{+}_{H}$ & $r^{-}_{H}$ & $\delta^{0.4}$   \\
            \hline
 0.58   & 1.80480   & 0.31720   & 1.48760        & 1.79094   & 0.40222   & 1.38872    \\
 0.68   & 1.72142   & 0.38435   & 1.33707        & 1.70460   & 0.47299   & 1.23161    \\
 0.78   & 1.61034   & 0.47826   & 1.13208       & 1.58784   & 0.57087   & 1.01697    \\
 0.88   & 1.45062   & 0.62166   & 0.82896       & 1.41262   & 0.72557   & 0.68705    \\
$a_{E}$ & 1.02901   & 1.02901   & 0.0           & 1.06144   & 1.06144   &0.0 \\
            \hline \hline
        \end{tabular}
    \end{center}
\end{table}
\begin{table}
    \begin{center}
        \caption{The event horizon ($r^{+}_{H}$) and the Cauchy horizon ($r^{-}_{H}$) of the black hole and their difference $\delta^{g}=r^{+}_{H}-r^{-}_{H}$ for $\alpha=\beta=0$ for different values of spin $a$, with constant $g$.}\label{table2}
        \begin{tabular}{l l l l | l l l}
            \hline \hline
            &\multicolumn{3}{c}{$g=0.3$}  &  \multicolumn{3}{c}{$g=0.4$}\\
            \hline
$a$ & $r^{+}_{H}$ & $r^{-}_{H}$ & $\delta^{0.3}$ & $r^{+}_{H}$ & $r^{-}_{H}$ & $\delta^{0.4}$   \\
            \hline
 0.58   & 1.80442   & 0.36513   & 1.43929        & 1.78999   & 0.45591   & 1.33408    \\
 0.68   & 1.72073   & 0.43301   & 1.28772        & 1.70287   & 0.53124   & 1.17163    \\
 0.78   & 1.60900   & 0.52439   & 1.08461        & 1.58430   & 0.63159   & 0.95271    \\
 0.88   & 1.44727   & 0.66341   & 0.78386        & 1.40249   & 0.79004   & 0.61245    \\
$a_{E}$ & 1.04466   & 1.04466   & 0.0            & 1.08796   & 1.08796   & 0.0 \\
            \hline \hline
        \end{tabular}
    \end{center}
\end{table}
To see the non trivial effect of the parameters $\alpha$ and $\beta$ on the horizon structure, we must choose
$\theta \neq \pi/2$.  The largest possible root of Eq.~(\ref{eh})
gives the location of the event horizon. We have studied the horizon
properties for nonzero values of $a$ and $g$ (cf.
Fig.~\ref{fig1}-\ref{fig3} and Table~\ref{table1},\ref{table2}) by numerically
solving Eq.~(\ref{eh}). We have demonstrated that for a
given value of $g$, there exist an extremal value of $a=a_{E}$ and
$r=r^{E}_{H}$ such that for $a<a_{E}$, Eq.~(\ref{eh}) admits two
positive roots and no root at $a>a_{E}$ (see
Fig.~\ref{fig1}-\ref{fig3}). It turns out that for $\alpha=1$,
$\beta=2$, $g=0.3$, we have $a_{E}=0.980078951651$ and $r^{E}_{H}=1.02901$.
Similarly for $\alpha=\beta=0$, $g=0.3$, one gets $a_{E}=0.9745094360075$ and
$r^{E}_{H}=1.04466$ (see Table~\ref{table1}, \ref{table2}). It can
be seen from the Tables~\ref{table1} and \ref{table2} that when
the value of $a$ increases, the radius of event horizon ($r^{+}_{H}$)
decreases and that of Cauchy horizon ($r^{-}_{H}$) increases. The difference
($\delta^{g}$) of radii of the horizons is listed in the
Tables~\ref{table1} and \ref{table2}. Interestingly, it turns
out that the $\delta^{g}$ decreases with increase in $a$ and it vanish
in the case of the extremal black hole. On the other hand, when the value of $g$
increases, the $\delta^{g}$ decreases. Indeed, the parameter
$\delta^{g}$ can be related to the area of ergoregion and the
ergoregion shrinks as the parameter $\delta^{g}$ decreases. Hence,
the parameter $g$ may play significant role in the energy extraction
process from the black holes. The Figs.~\ref{fig1} and \ref{fig2}
shows that there exists a set of values of the parameters for which the
black hole (\ref{hay}) has two horizons or we have a regular black
hole with the Cauchy and the event horizons. Further, one can find
values of parameters for which these two horizons coincide which correspond to an extremal black hole.

Next, we investigate the structure and the location of the ergosurface or the static limit surface ($r_{H}^{sls}$), which requires coefficient of $dt^2$ to be zero. Then it is clear from Eq.~(\ref{hay}) that static limit surface satisfies
\begin{eqnarray}\label{sls}
r^2(r^{3+\alpha}\Sigma^{-\alpha/2}+g^3r^{\beta}\Sigma^{-\beta/2})-2Mr^{4+\alpha}\Sigma^{-\alpha/2}\nonumber \\
+a^2 (r^{3+\alpha}\Sigma^{-\alpha/2}+g^3r^{\beta}\Sigma^{-\beta/2})\cos^2 \theta &=&0.\nonumber \\
\end{eqnarray}
The location of static limit surface is shown in the Fig.~\ref{fig4} for different values of $a$ and $g$. It is observed that for $g=0$, Eq.~(\ref{eh}) and (\ref{sls}) are exactly same as the Kerr black hole \cite{Kerr:1963ud}. The ergosphere is the region between the static limit surface and the event horizon. It lies outside the black hole and it is possible to enter an ergosphere and leave again, a object moves in the direction of spin of the black hole. Interestingly, the area of ergoregion decreases with increases in $a$ as well as with increase in $g$. Thus, the event horizon of the rotating Hayward's regular metric (\ref{hay}) is located at $r=r_{H}^{E}$, where $\Delta=0$, and it is rotating with angular velocity $\Omega_{H}$. Whereas the static limit surface is located at $r=r_{H}^{sls}$, when $g_{tt}=0$. Further, one has an extremal black holes, when $\Delta=0$ has a double root, i.e., when the two horizons coincides. The ergoregion is given by $r_{H}^{E}<r<r_{H}^{sls}$, where the Killing vector $\frac{\partial}{\partial t}$ is spacelike. When $\Delta=0$ has no root, i.e., no horizon exists, one gets a naked singularity or no black hole (cf. Fig.~\ref{fig4} and \ref{fig5} ).

\section{Particles orbits}
In this section, we would like to study the equations of motion of a particle with rest mass $m_{0}$ falling in the background of the rotating Hayward's regular black hole. Henceforth, we shall restrict our discussion to the case of equatorial plane ($\theta=\pi/2 $), which simplifies the mass function (\ref{m}) to
\begin{equation}\label{m1}
m= \frac{M}{1+(g/r)^3}.
\end{equation}
It is easy to see that the mass function (\ref{m1}) can be also obtained from (\ref{m}) for $\alpha=\beta=0$.
Further, the equations $E=-p_{t}$ and $L=p_{\phi}$, lead to
\begin{equation}\label{E}
E = -(g_{tt} \dot{t}+g_{t\phi}\dot{\phi}),
\end{equation}
\begin{equation}\label{L}
L = g_{\phi \phi} \dot{\phi}+g_{t\phi}\dot{t},
\end{equation}
which trivially solves to
\begin{equation}\label{eqm1}
 \Sigma \frac{d t}{d \tau} =  -a (aE - L) + \left(r^2 + a^2\right) \frac{T}{\Delta},
\end{equation}
\begin{equation}\label{eqm3}
\Sigma \frac{d \phi}{d \tau} = -(aE -L) + \frac{a T}{\Delta}.
\end{equation}
The radial part of the equation of motion of a particle in the rotating Hayward's spacetime can be analysed by the Hamilton-Jacobi separation method. The Hamilton for the geodesic motion is given by
\begin{equation}\label{ham}
H = \frac{1}{2} g^{\mu\nu} P_{\mu} P_{\nu},
\end{equation}
where $P_{\mu}$ is the momentum. If $S=S(\lambda,x^{\alpha})$ be the action which is a function of the parameter $\lambda$ and coordinate $x^{\alpha}$. Then, the corresponding Hamilton-Jacobi equation
\begin{equation}\label{hje}
\frac{\partial S}{\partial \tau} = -\frac{1}{2} g^{\mu\nu} \frac{\partial S}{\partial x^{\mu}} \frac{\partial S}{\partial x^{\nu}},
\end{equation}
where $\tau$ is an affine parameter along the geodesics and $S$ is Jacobi action which is given by
\begin{equation}\label{hja}
S = \frac{1}{2} m_0^2 \tau -Et + L \phi + S_{r}(r),
\end{equation}
where $ S_{r} $ is a function of $ r $. The constants $ m_0 $, $ E $, and $ L $ correspond to rest mass, conserved energy and angular momentum of the particle, respectively. They are related via $m_{0}^2= -p_{\mu}p^{\mu}$, $E=-p_{t}$, and $L=p_{\phi}$. In addition to $E$ and $L$, there is another conserved quantity, namely Carter constant $K$, which is related to total angular momentum. Inserting Eq.~(\ref{hja}) into the Eq.~(\ref{hje}) and separating the coefficients of $r$ equal to the Carter constant $K$, we get the following equation
\begin{equation}\label{eqm4}
\Sigma \frac{d r}{d \tau} = \pm \sqrt{T^2 -\Delta \left[m_{0}^2 r^2  + (L-a E)^2 + K\right]},
\end{equation}
where $ T = E (r^2 + a^2) -La $. For a particle moving in the equatorial plane and to remain in the equatorial plane, the Carter constant $K=0$ \cite{Bardeen:1972fi}. Obviously, one recovers the equations of motion of the Kerr black hole when $g \rightarrow 0$ \cite{Banados:2009pr}.

\begin{figure}[tbp]
\centering 
\includegraphics[width=.45\textwidth]{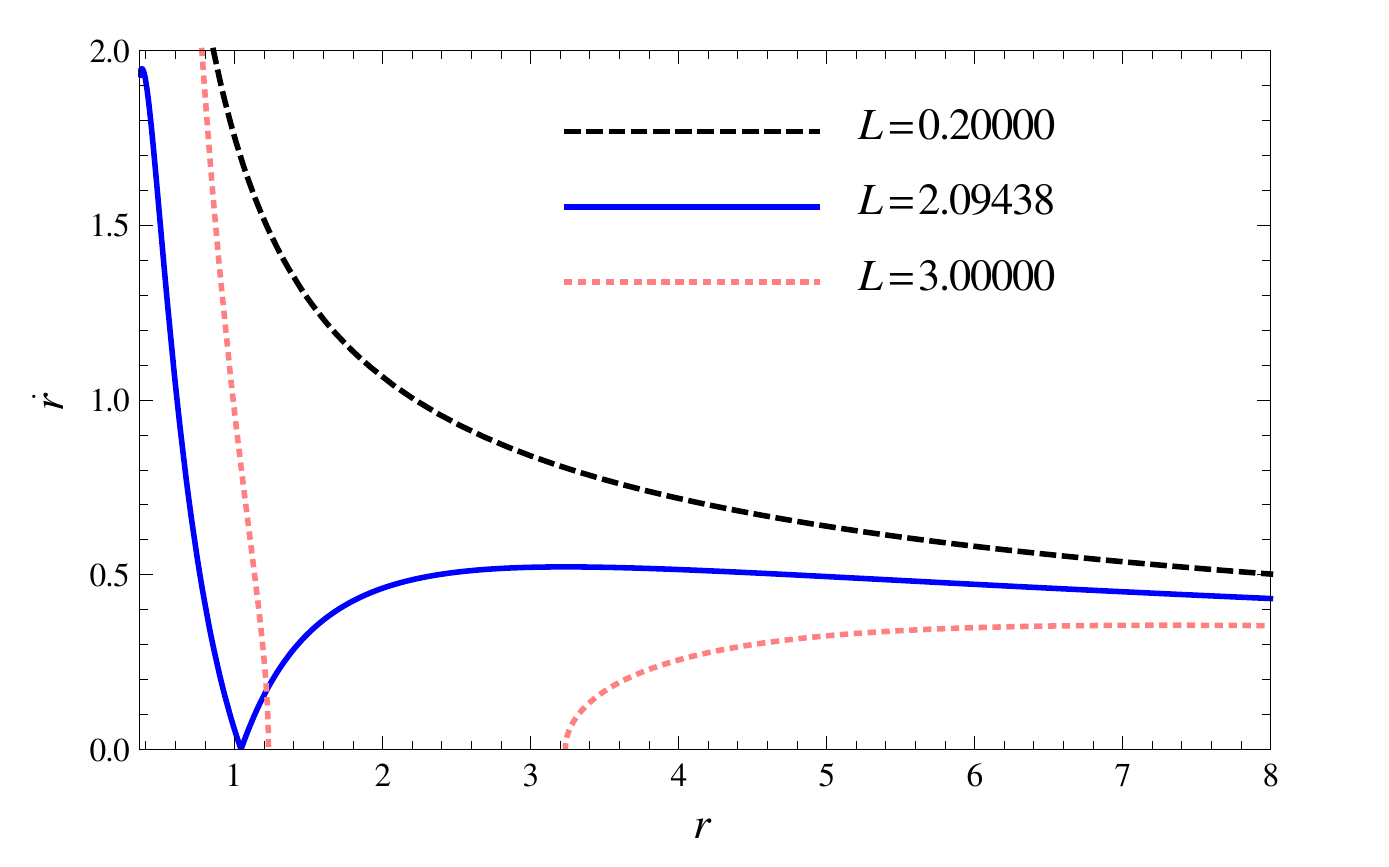}
\includegraphics[width=.45\textwidth]{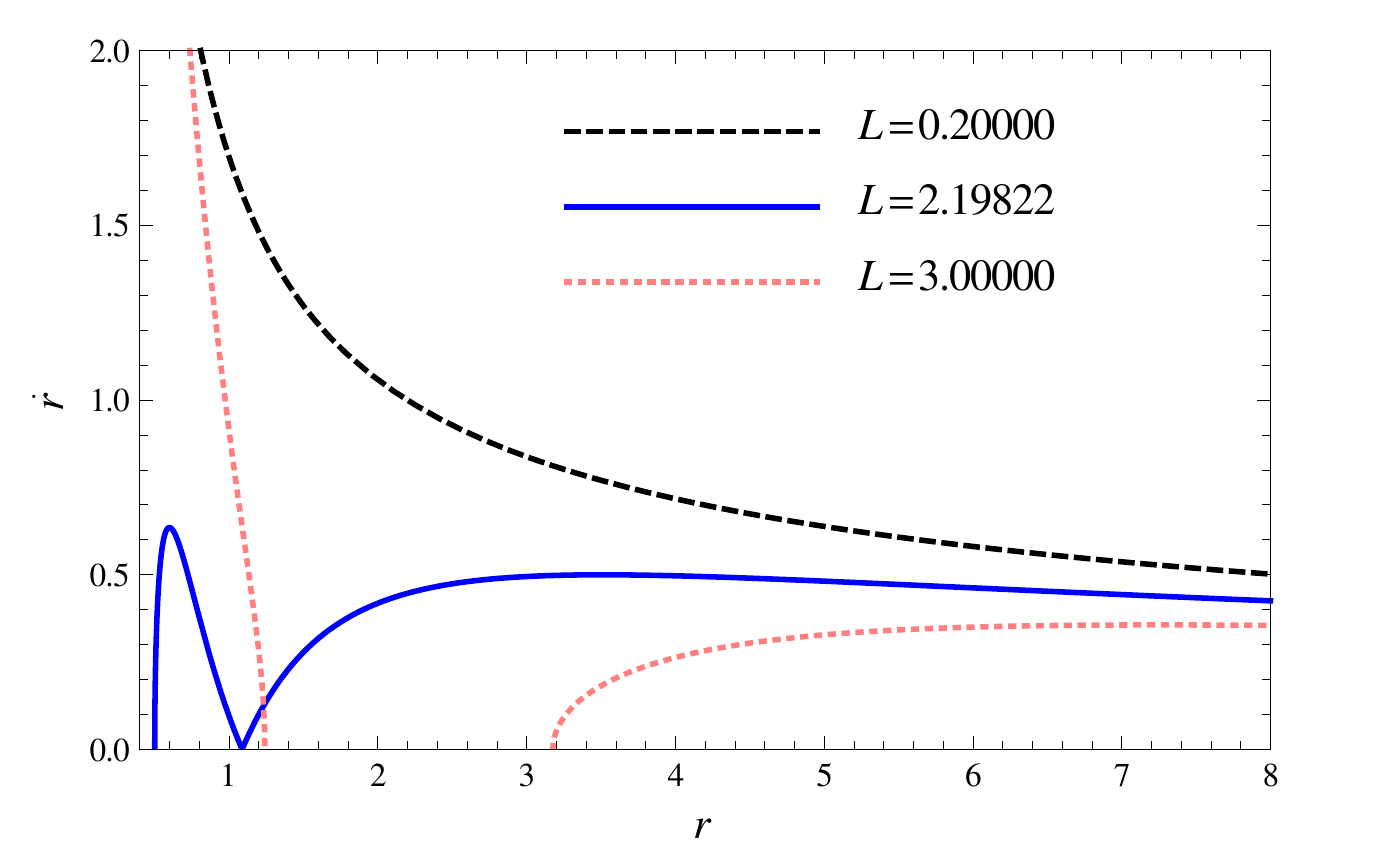}
\caption{\label{fig6} Plots showing the behavior of $\dot{r}$ vs $r$ for different values of angular momentum $L$. (Left) For $a=a_{E}=0.9745094360075$ and $g=0.3$. (Right) For $a=a_{E}=0.9429970792861$ and $g=0.4$.}
\end{figure}

\begin{figure}[tbp]
\centering 
\includegraphics[width=.45\textwidth]{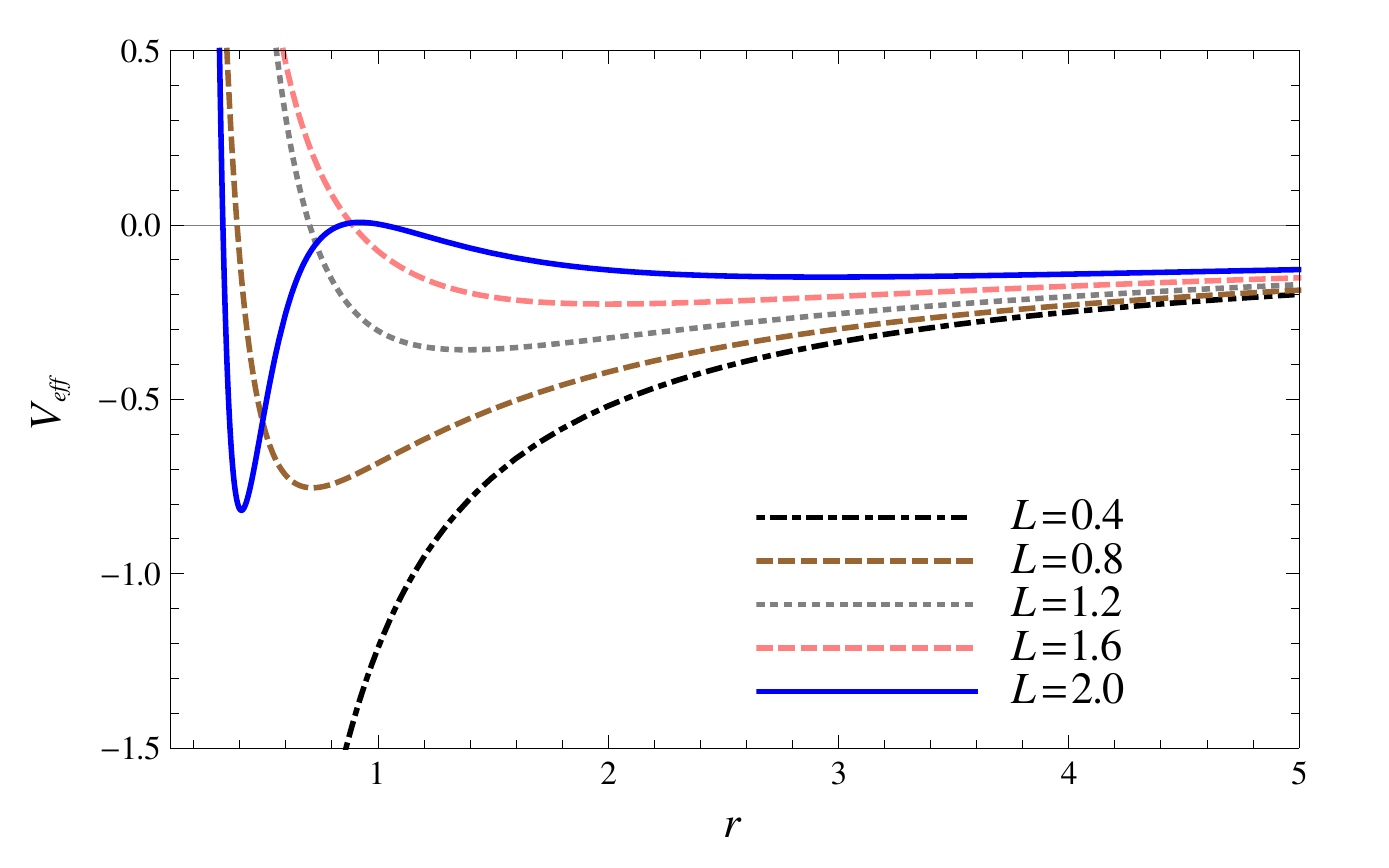}
\includegraphics[width=.45\textwidth]{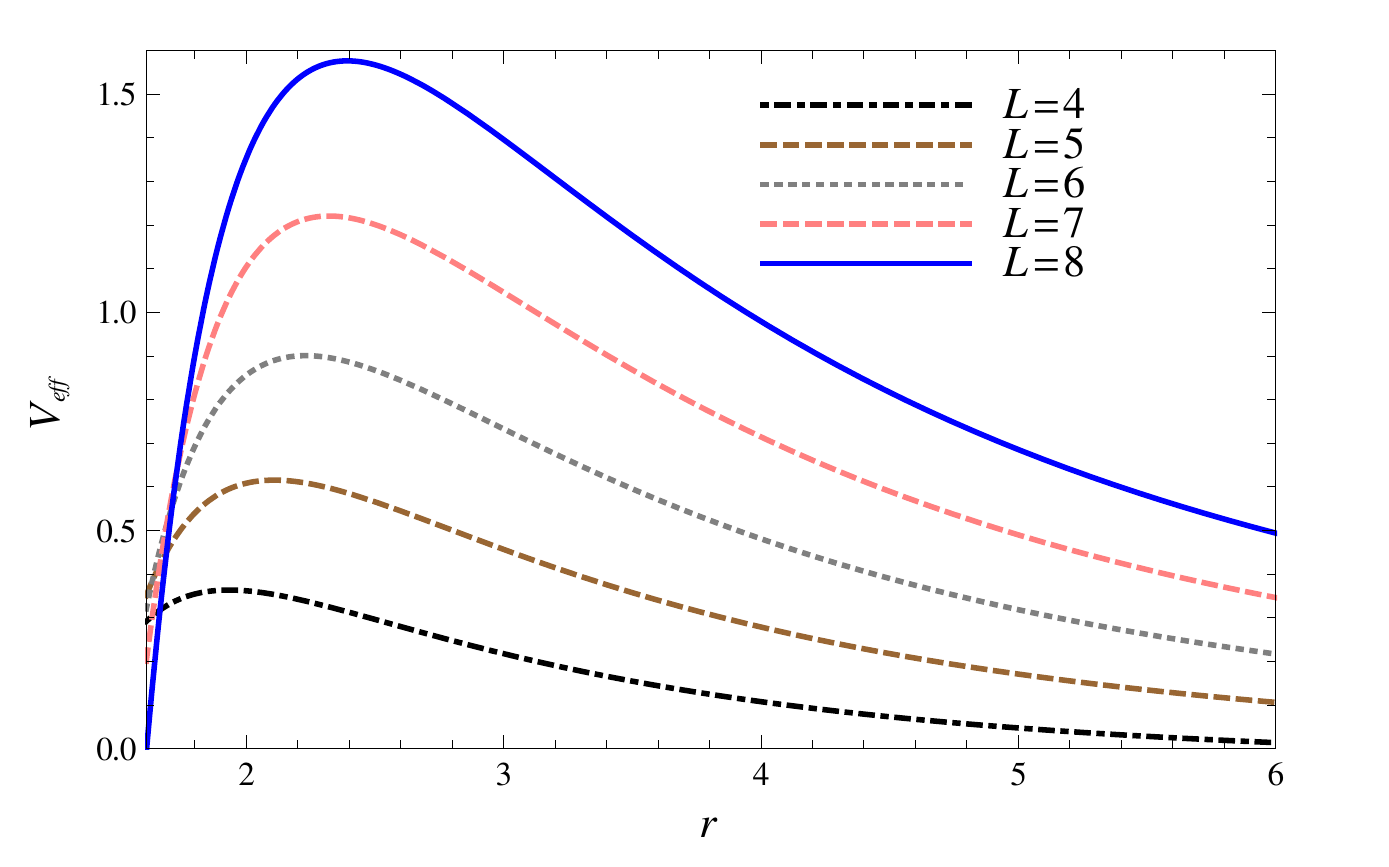}
\caption{\label{fig7} Plots showing the behavior of $V_{eff}$ vs $r$ for $a=a_{E}=0.9745094360075$ and $g=0.3$. (Left) For when $L$ lies in the range. (Right) For when $L$ lies outside the range.}
\end{figure}

To determine the range of the angular momentum of the particles, we must calculate the effective potential. The radial equation for the timelike particle moving along the geodesic in the equatorial plane is described by
\begin{equation}
\frac{1}{2} \dot{r}^2+ V_{eff}=0,
\end{equation}
with the effective potential
\begin{equation}
 V_{eff} =  -\frac{[E (r^2 + a^2) -La]^2 -\Delta [m_{0}^2 r^2  + (L-a E)^2]}{2r^4}.
\end{equation}
The condition of circular orbit of the particles is given by
\begin{equation}\label{lim}
V_{eff}=0, \;\;\; \mbox{and} \;\;\; \frac{dV_{eff}}{dr}=0.
\end{equation}
\begin{table}
    \begin{center}
        \caption{The range for the angular momentum $L$, and value of $a=a_{E}$, $r=r_{H}^E$ for different $g$ of the extremal rotating Hayward's regular black hole.}\label{table3}
        \begin{tabular}{l l | l | l l}
            \hline \hline
$g$   & $a_{E}$                & $r^{E}_{H}$  & $L_{2}$(min)  & $L_{1}$(max) \\
\hline
0.2   & 0.9921503552200    & 1.01501      &-4.82270     & 2.03055  \\
0.3   & 0.9745094360075    & 1.04466      &-4.80977    & 2.09438  \\
0.4   & 0.9429970792861    & 1.08796      &-4.78643    & 2.19822  \\
0.5   & 0.8961057957904    & 1.13802      &-4.75113    & 2.34136  \\
0.6   & 0.8327653930500    & 1.18888      &-4.70236    & 2.52988  \\
\hline \hline
        \end{tabular}
     \end{center}
\end{table}
\begin{table}
    \begin{center}
        \caption{The range for the angular momentum $L$, and values of $a$, $r=r_{H}^{-}, r_{H}^{+}$ for different $g$ of the non-extremal rotating Hayward's regular black hole.}\label{table4}
        \begin{tabular}{l l | l l | l l}
            \hline \hline
 $g$   & $a$    & $r^{-}_{H}$   & $r^{+}_{H}$  & $L_{4}$(min)  & $L_{3}$(max)\\
\hline
 0.2   & 0.96   & 0.77195   & 1.26151  & -4.79983   & 2.37942 \\
 0.3   & 0.88   & 0.66341   & 1.44727  & -4.74164   & 2.66755 \\
 0.4   & 0.78   & 0.63159   & 1.58430  & -4.66671   & 2.90664 \\
 0.5   & 0.72   & 0.68597   & 1.62411  & -4.61962   & 3.01039 \\
 0.6   & 0.68   & 0.77888   & 1.61674  & -4.58625   & 3.05693 \\
\hline \hline
        \end{tabular}
\end{center}
\end{table}
Since geodesics are timelike, i.e., $dt/d\tau \geq 0$, then Eq.~(\ref{eqm1}) leads to
\begin{equation}
\frac{1}{r^2} \Big[-a (aE - L) + \left(r^2 + a^2\right) \frac{T}{\Delta}\Big] \geq 0,
\end{equation}
the above condition, as $r \rightarrow r^{E}_{H}$, reduces to
\begin{eqnarray}\label{cnd}
E -\Omega_{H}L &\geq & 0,\nonumber \\
 \Omega_{H}&=&\frac{a}{2mr^{E}_{H}} = \frac{a}{(r^{E}_{H})^2+a^2},
\end{eqnarray}
where $\Omega_{H}$ is the angular velocity of the black hole on the horizon. The limiting values of the angular momentum of freely falling particles are calculated using Eq.~(\ref{lim}) for both the extremal and the non-extremal rotating Hayward's regular black hole, which are listed in Tables~\ref{table3} and \ref{table4}. The critical angular momentum of the particle is given by $ L_{c} \equiv E/ \Omega_{H} $, and from (\ref{cnd}), $L\leq L_{c}$. The values of critical angular momentum are given in the Table~\ref{table3} for different combinations of spin $a$ and constant $g$. In Fig.~\ref{fig6}, we plot $\dot{r}$ vs $r$ for the different values of $L$, $a$ and $g$. It is shown that if the angular momentum of the particle is larger than the critical angular momentum, i.e., if $L>L_{c}$, then the geodesics never fall into the black hole. On the other hand, if the angular momentum is smaller than the critical angular momentum ($L<L_{c}$), then the geodesics always fall into the black hole and if both are equal ($L=L_{c}$), then the geodesics fall into the black hole exactly at the event horizon. We plot the effective potential in Fig.~\ref{fig7} choosing different values of angular momentum $L$. If the angular momentum of the particles lie in the range, then the effective potential is negative and particles are bounded. If the angular momentum of the particle lie outside the range, then the effective potential is always positive.

\section{Center-of-mass energy in the rotating Hayward's regular black hole}
\begin{figure}[tbp]
\centering 
\includegraphics[width=.45\textwidth]{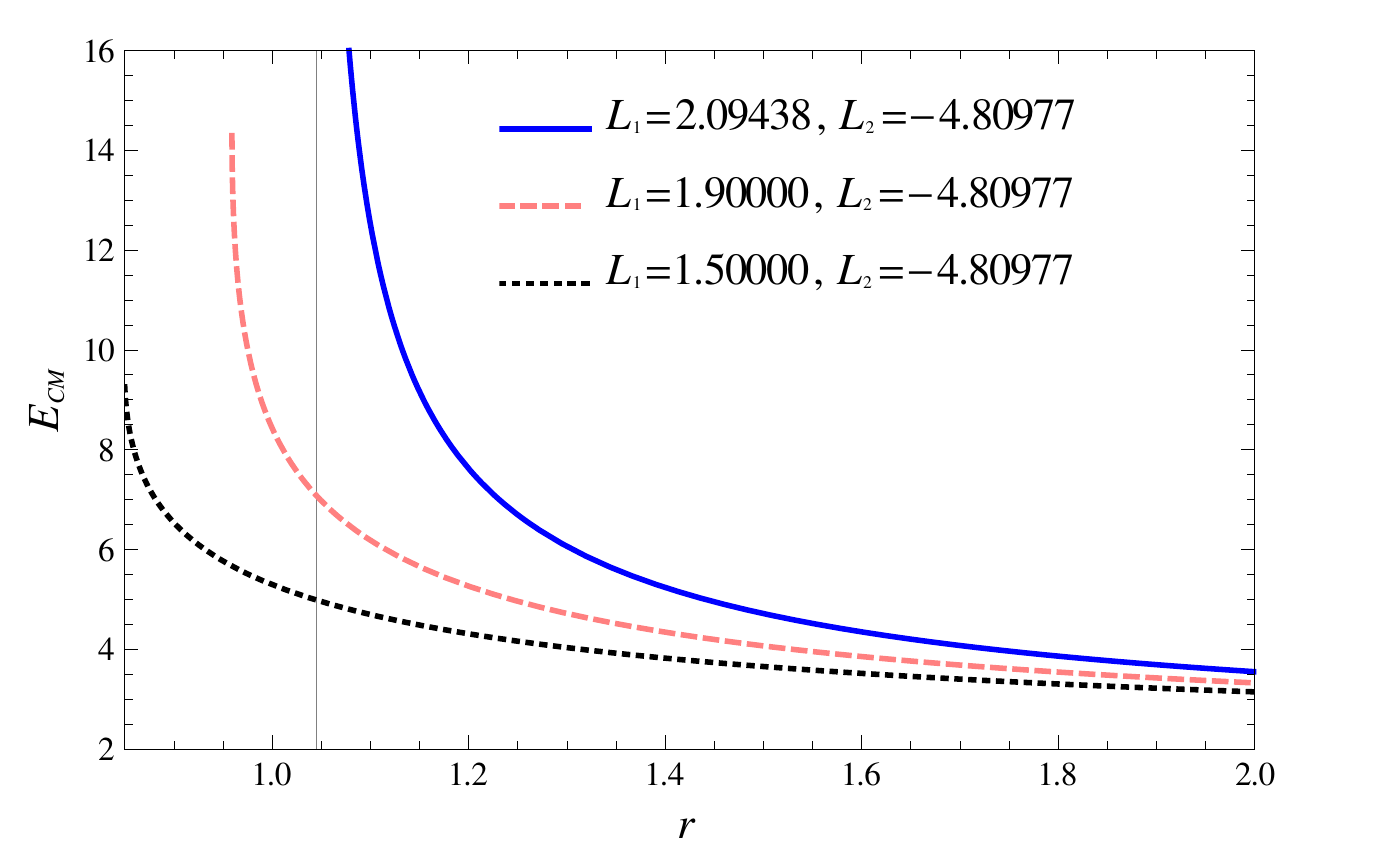}
\includegraphics[width=.45\textwidth]{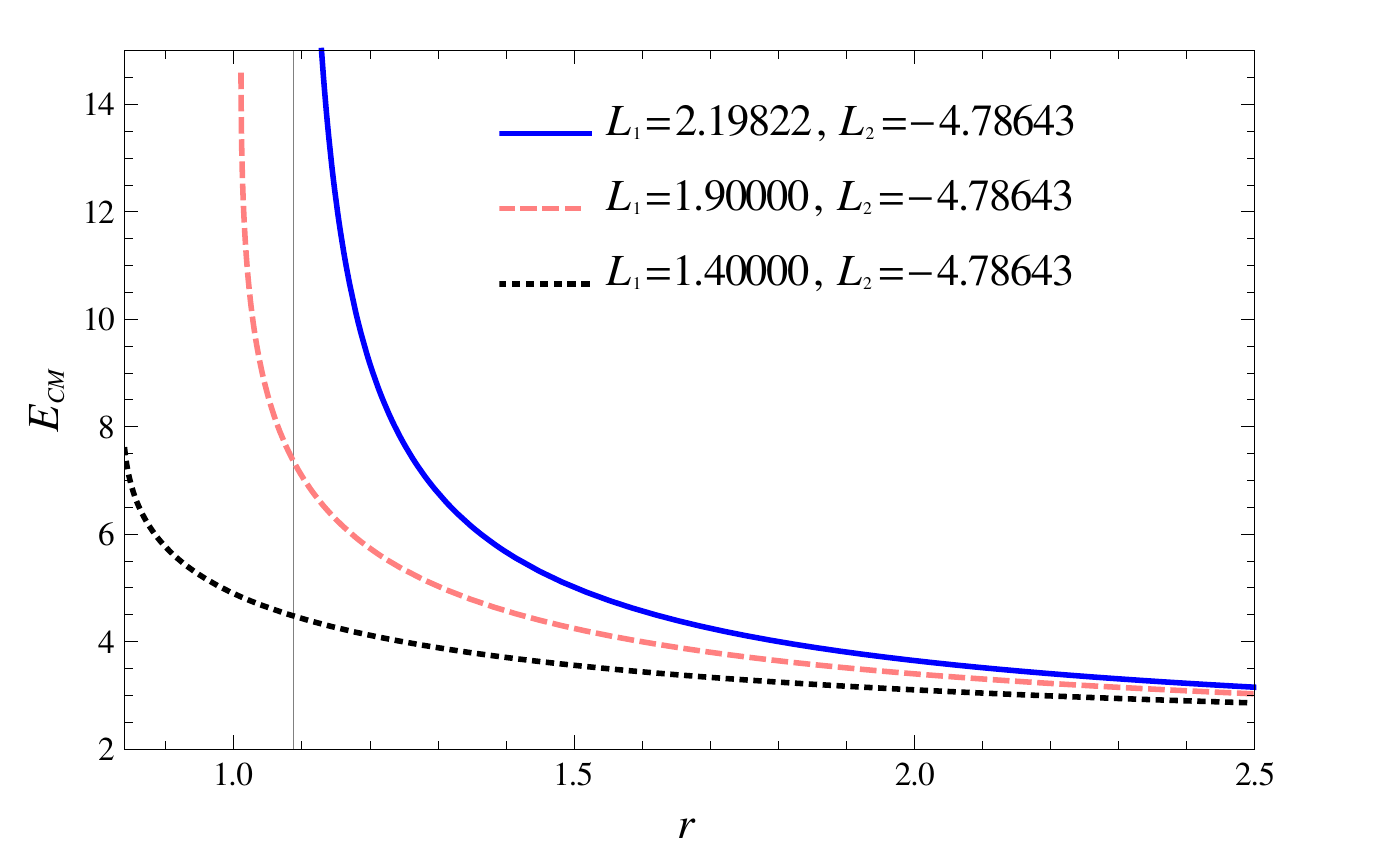}
\caption{\label{fig8} Plots showing the behavior of $E_{CM}$ vs $r$ for extremal black hole. (Left) For $a=a_{E}=0.9745094360075$ and $g=0.3$. (Right) For $a=a_{E}=0.9429970792861$ and $g=0.4$ where the vertical line denotes the event horizon.}
\end{figure}

In the last section, we calculate the range for the angular momentum for which a particle can reach the horizon, i.e., if the angular momentum lies in the range, the collision is possible near the horizon of the rotating Hayward's regular black hole. Next, we study the $E_{CM}$ of two colliding particles moving in the equatorial plane of the rotating Hayward's regular black hole. Let us consider colliding particles have same rest mass $m_{1}= m_{2}=m_{0}$, and they are coming from rest at infinity with $E_{1}/m_{0}=E_{2}/m_{0}=1$, approaching the black hole with different angular momenta $ L_{1}$, $ L_{2}$ and collide at some radius $ r $. We wish to compute the collision energy of the particles in center-of-mass frame and explicitly bring out the effect of the parameter $g$ on the BSW mechanism. We observe that two particles with the four-momentum
\begin{equation}\label{momentum}
P_{i}^{\mu}=m_{i}u_{i}^{\mu},
\end{equation}
where $u^{\mu}_{i}=dx^{\mu}_{i}/d\tau$ is the four-velocity of the particles $i$ ($i=1,2$). The $E_{CM}$ of two particles is given by
\begin{equation}\label{formula}
E_{CM}^2=-P_{i}^{\mu}P_{i\mu}.
\end{equation}
Inserting, Eq.~(\ref{momentum}) into Eq.~(\ref{formula}), with $m_{1}= m_{2}=m_{0}$, we obtain
\begin{equation}\label{eqlm}
\frac{E_{CM}^2}{2 m_{0}^2 } =  1-g_{\mu \nu} u^{\mu}_{(1)} u^{\nu}_{(2)}.
\end{equation}
On substituting the values of $g_{\mu \nu}$, $u^{\mu}_{(1)}$ and $u^{\mu}_{(2)}$, the expression for the $E_{CM}$ of two colliding particles has the following simple form:
\begin{eqnarray}\label{ecm}
\frac{E_{CM}^2}{2 m_{0}^2 } &=& \frac{1}{r(r^2-2mr+a^2)} \Big[2a^2(m+r)
\nonumber \\
&-& 2am(L_{1}+L_{2})
\nonumber \\
&-& L_{1} L_{2}(-2m+r) +2(-m+r)r^2
\nonumber \\
&-&\sqrt{2m(a-L_{1})^2 - L_{1}^2r + 2mr^2}
\nonumber \\ &&
\sqrt{2m(a-L_{2})^2 - L_{2}^2r + 2mr^2}\Big],
\end{eqnarray}
where $m$ is given by Eq.~(\ref{m1}). Obviously, the above result confirms that the parameter $g$ indeed has influence on the $E_{CM}$, and when the deviation parameter vanish $g=0$, the above equation reduces to same as obtained for the Kerr black hole \cite{Banados:2009pr}. We are interested to investigate the properties of the $E_{CM}$ as $r\rightarrow r_{H}^{E}$. We observe that at $r \rightarrow r^{E}_{H}$ the denominator of Eq.~(\ref{ecm}) is zero and so is the numerator. We apply l'Hospital's rule twice, then the value of the $E_{CM}$, as $r\rightarrow r_{H}^{E}$, becomes
\begin{eqnarray}\label{ecm2}
\frac{E_{CM}^2}{2 m_{0}^2 }(r\rightarrow r^{E}_{H}) &=& 3.7942+0.0812(L_{1}+L_{2})
\nonumber \\
&-& 0.0819L_{1}L_{2} +\frac{G^{2}_{1}(L_{2}-L_{c})}{8(L_{1}-L_{c})^3}
\nonumber \\
&+& \frac{G^{2}_{2}(L_{1}-L_{c})}{8(L_{2}-L_{c})^3}-\frac{G_{1}G_{2}}{4(L_{1}-L_{c})(L_{2}-L_{c})}
\nonumber \\
&-&\frac{H_{1}(L_{2}-L_{c})}{4(L_{1}-L_{c})}-\frac{H_{2}(L_{1}-L_{c})}{4(L_{2}-L_{c})},
\end{eqnarray}
where $a=a_{E}=0.9745094360075$, $r^{E}_{H}=1.04466$, $g=0.3$, $M=1$, $G_{i}=4.1189-0.0883L_{i}-0.9554L^{2}_{i}$, $H_{i}=3.8008+0.3443L_{i}-0.1735L^{2}_{i}$ ($i=1,2$), and $L_{c}=E/\Omega_{H}=2.09438$. Eq.~(\ref{ecm2}), gives the limiting values of the $E_{CM}$ at the event horizon with critical angular momentum $L_{c}$. Obviously, the $E_{CM}$ diverges when $L_{1}=L_{c}$ or $L_{2}=L_{c}$ and the $E_{CM}$ is finite for the other generic values of $L_{1}$ and $L_{2}$. Hence, we can say that an extremal rotating Hayward's regular black hole can act as a particle accelerator to an infinitely high energy and may provide an effective framework for the Planck scale physics. However, to get infinite $E_{CM}$ the particles should  approach the black hole with angular momentum in the required range, which is reflected in Table~\ref{table3}. Further, Fig.~\ref{fig8} depicts the variation of the $E_{CM}$ vs $r$ for different values of $L_{1}$ and $L_{2}$ with fixed values of $a$ and $g$, it is clear that the $E_{CM}$ blows up at the horizon when either $L_{1}$ or $L_{2} =L_{c}$. It is clear that $L_{c}$ is in the range for which the particle can reach the horizon of the black hole.

\begin{figure}[tbp]
\centering 
\includegraphics[width=.45\textwidth]{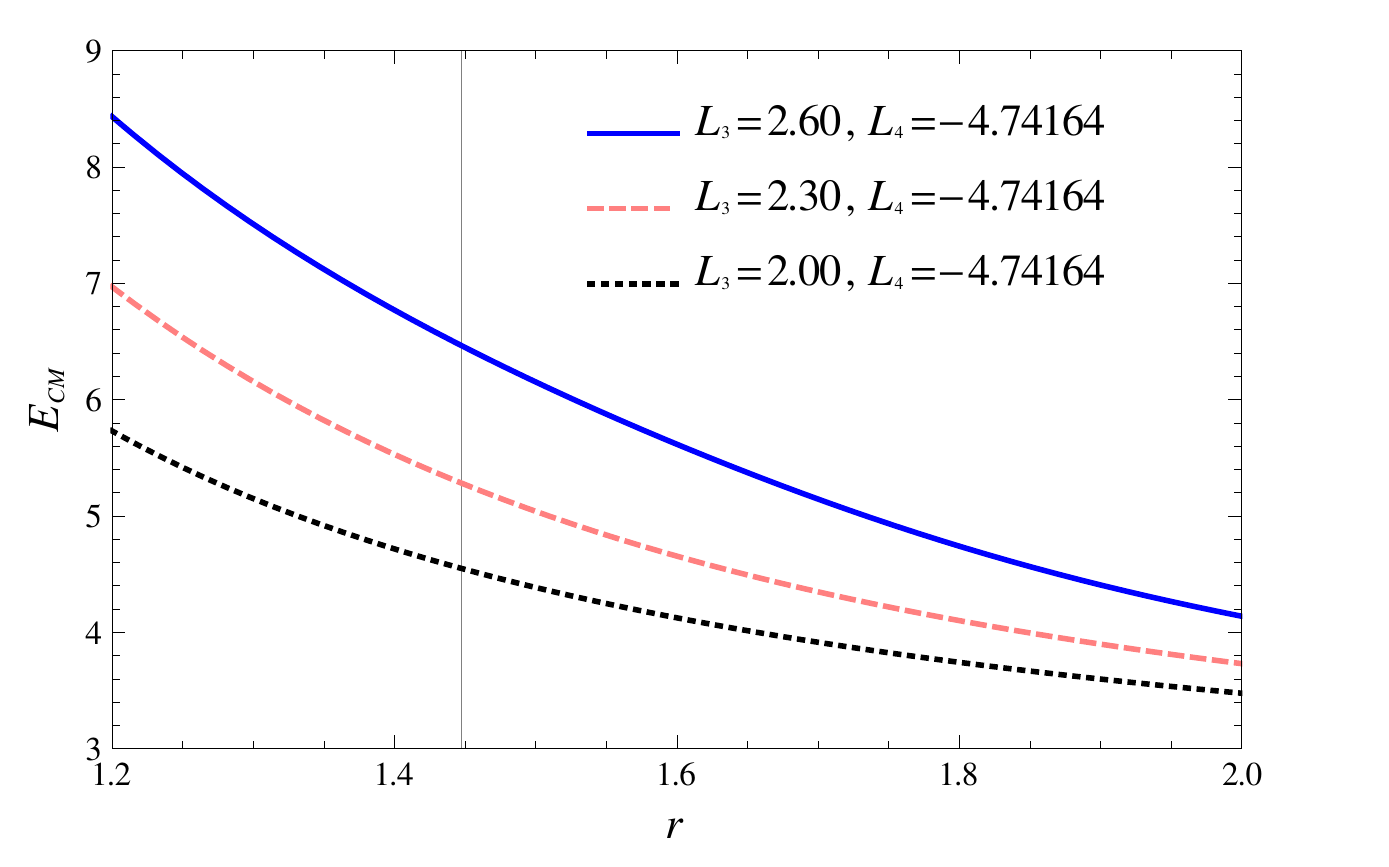}
\includegraphics[width=.45\textwidth]{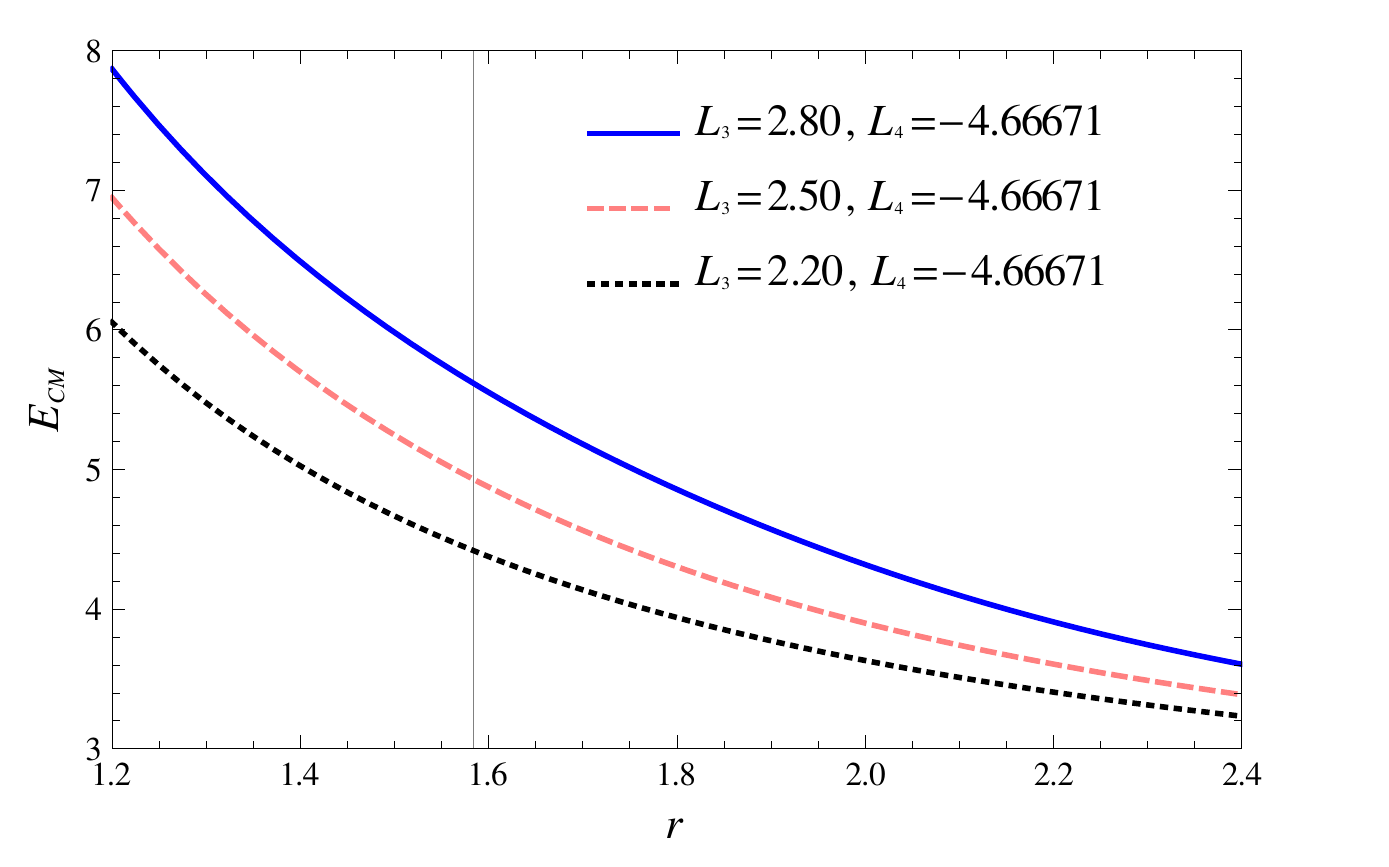}
\caption{\label{fig9} Plots showing the behavior of $E_{CM}$ vs $r$ for non-extremal black hole. (Left) For $a=0.88$ and $g=0.3$. (Right) For $a=0.78$ and $g=0.4$ where the vertical line denotes the event horizon.}
\end{figure}

\begin{figure}[tbp]
\centering 
\includegraphics[width=.45\textwidth]{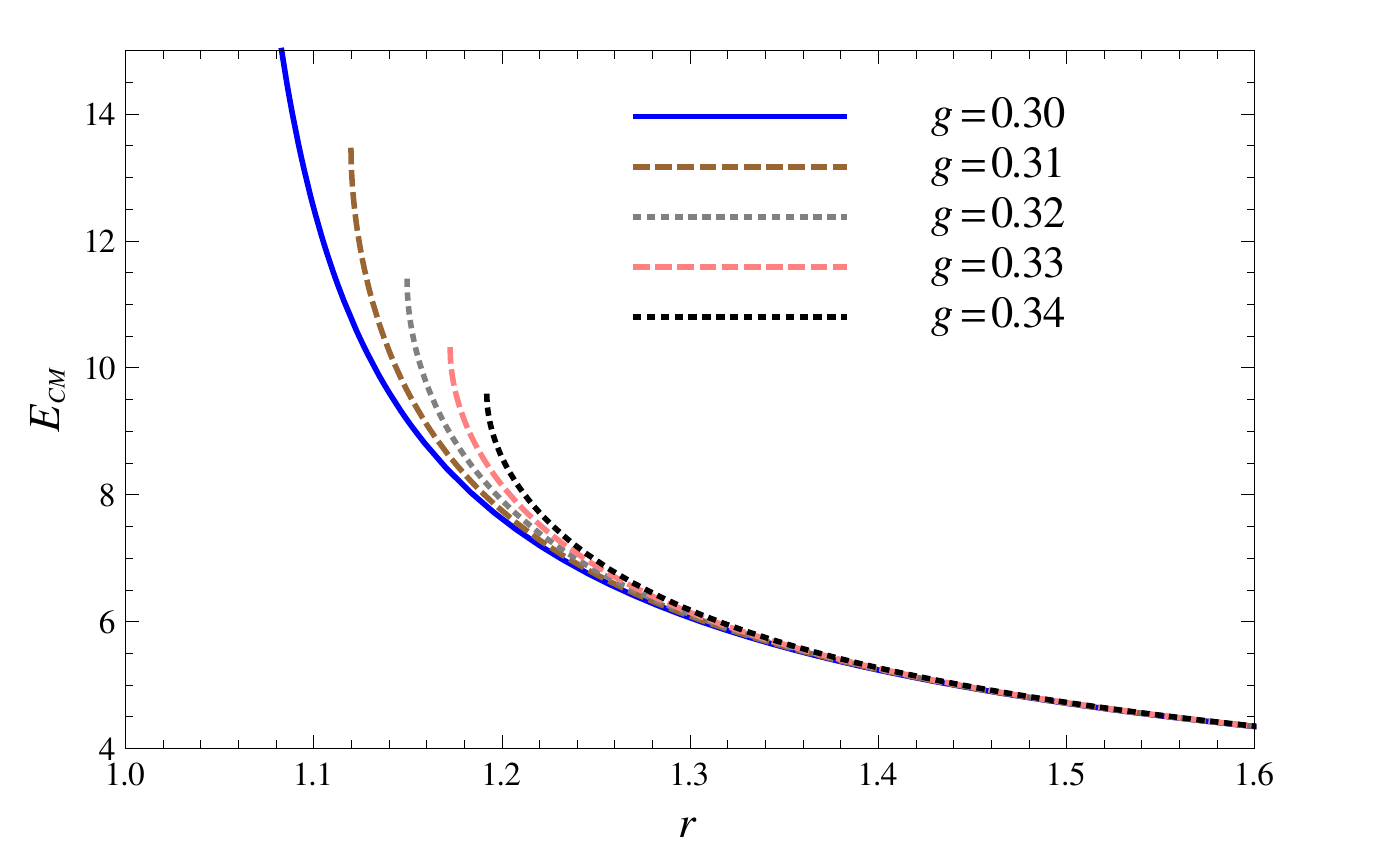}
\includegraphics[width=.45\textwidth]{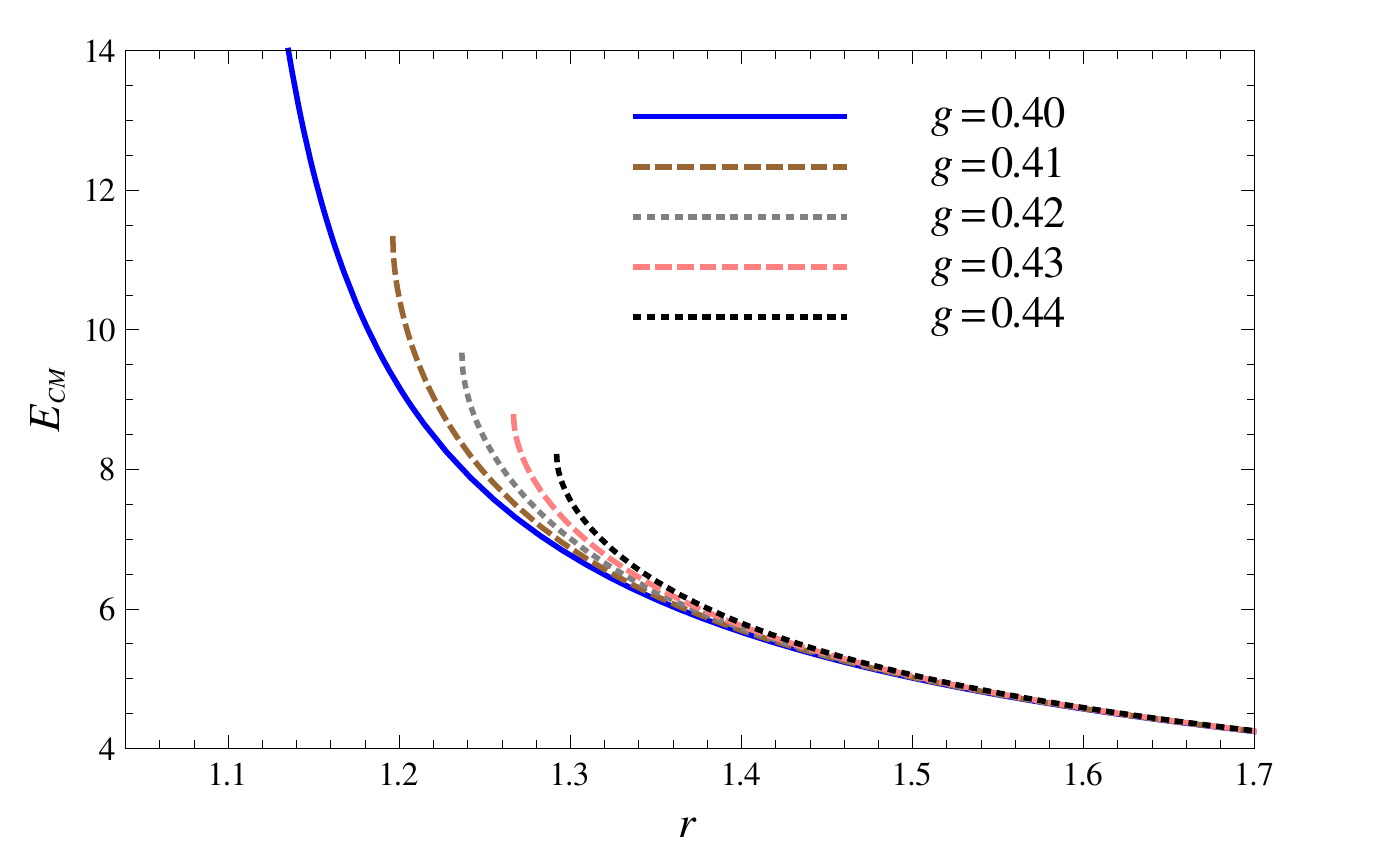}
\caption{\label{fig10} Plots showing the behavior of $E_{CM}$ vs $r$ for different values of $g$. (Left) For $a=a_{E}=0.9745094360075$. (Right) For $a=a_{E}=0.9429970792861$.}
\end{figure}

\paragraph{Particle collision near non-extremal black hole}
Next, we want to study the properties of the $E_{CM}$ as $r$ tends to the event horizon $r^{+}_{H}$ in the case of the non-extremal rotating Hayward's regular black hole. At $r\rightarrow r_{H}^{+}$, both numerator and denominator of the Eq.~(\ref{ecm}) vanish. So, we apply l'Hospital's rule to find the near horizon $E_{CM}$ for the non-extremal black hole. The $E_{CM}$, as $r\rightarrow r_{H}^{+}$, is given as
\begin{eqnarray}\label{ecm3}
\frac{E_{CM}^2}{2 m_{0}^2 }(r\rightarrow r^{+}_{H}) &=& \frac{1}{0.6657(L_{3}-L'_{c})(L_{4}-L'_{c})}\Big[\nonumber \\ &&
14.1523+L_{4}(1.2332L_{4}-4.3409) \nonumber \\
&+&L_{3}(1.2332L_{3}-1.1349L_{4}-4.3409)\Big],\nonumber \\
\end{eqnarray}
where $L'_{c}=E/\Omega_{H}=3.2602$. From Eq.~(\ref{ecm3}), it seems that one can get infinite $E_{CM}$ if either $L_{3}$ or $L_{4}=L'_{c}$. However, we have to guarantee that the particles with angular momentum $L'_{c}$ reaches the horizon or in other words $L'_{c}$ should be in the range of angular momentum with which the particles can reach the horizon and collision is possible. It can be seen from Table~\ref{table4}, the range of the angular momentum of the particles, for $g=0.3$, and $a=0.88$, is $L_{4} <L <L_{3}$. It turns out that the value of $L'_{c}$ does not lie in the range and $L'_{c}>L_{3}$, which means that, in non-extremal rotating Hayward's regular black hole, the particles with angular momentum $L=L'_{c}$ could not fall into the black hole. Hence, the $E_{CM}$ for non-extremal rotating Hayward's regular black hole has a finite upper limit. For non-extremal black hole, the behavior of the $E_{CM}$ vs $r$ can be seen from Fig.~\ref{fig9} for different values of the parameters $a$ and $g$. We can also observe, from the Fig.~\ref{fig10}, the $E_{CM}$ is increases with $g$.

\section{Conclusion}
The celebrated singularity theorems predict the formation of singularities in classical general relativity. However, it is widely believed that spacetime singularities do not exist in nature, but that they represent a limitation or creation of the classical general theory of relativity. As we do not yet have any well defined theory of quantum gravity, hence more attention is given for phenomenological approaches  to somehow solve these singularity problem in classical general relativity and to study possible implications. So an important line of research to understand the inside of a black hole is tantamount to investigate classical black holes and their consequences, with regular, i.e., nonsingular, properties. In view of this, we have examined the features of horizons by the stationary, rotating Hayward's regular black hole and explicitly bring out the effect of the deviation parameter $g$. It turns out that for each $g$, for proper choice of parameters $\alpha$, $\beta$, $M$, and $\theta$, we can find critical value $a=a_{E}$, which corresponds to an extremal black hole with degenerate horizons, i.e., where two horizons coincides. However, when $a<a_{E}$, we have a regular black hole with Cauchy and event horizon. It turns out that the horizon structure of the rotating Hayward's regular black hole is complicated as compared to the Kerr black hole. Thus, the extremal regular black hole depends on the value of $g$. Further, we have adapted the original BSW mechanism suitable for the rotating Hayward's regular black hole, which has very complicated horizon structure as compared to the Kerr black hole. Then, we study the collision of two particles of equal rest masses falling freely from rest at infinity into the equatorial plane of an extremal rotating Hayward's regular black hole to calculate the $E_{CM}$, for various values of $g$, which are infinite if one of the colliding particles has the critical angular momentum in the required range. On the other hand, the $E_{CM}$ has always finite upper limit for the non-extremal black hole. Thus, the BSW mechanism depends both on the rotation parameter $a$ as well as on the deviation parameter $g$. For the non-extremal black hole, we have also seen the effect of $g$ on the $E_{CM}$ demonstrate a increase in the value of the $E_{CM}$ with an increase in the value of $g$. We have performed our calculations numerically as it is difficult to get the analytical solution and found that the results are different from that of the Kerr case. In particular, our results, in the limit $g \rightarrow 0$, reduced exactly  to  \emph{vis-$\grave{a}$-vis}  the Kerr black hole results.

\acknowledgments

Authors would like to thank IUCAA, pune for hospitality while part of this work was done.  M.A. acknowledges the University Grant Commission, India, for financial support through the Maulana Azad National Fellowship For Minority Students scheme (Grant No.~F1-17.1/2012-13/MANF-2012-13-MUS-RAJ-8679). S.G.G. would like to thank SERB-DST for Research Project Grant NO SB/S2/HEP-008/2014.

\end{document}